\documentclass{article}

\usepackage{a4wide}
\usepackage{amsmath,amssymb,amscd,amsfonts}

\usepackage{mathrsfs,latexsym}

\newcommand{\II}{{\boldsymbol{1}}}

\newcommand{\CC}{{\mathbb C}}
\newcommand{\RR}{{\mathbb R}}

\newcommand{\TT}{{\mathbb T}}
\newcommand{\ZZ}{{\mathbb Z}}

\newcommand{\CoinX}[1]{C_0^\infty({#1})}

\newtheorem{Thm}{Theorem}[section]
\newtheorem{Def}[Thm]{Definition}
\newtheorem{Lem}[Thm]{Lemma}
\newtheorem{Prop}[Thm]{Proposition}

\newcommand{\DD}{{\mathscr D}}
\newcommand{\EE}{{\mathscr E}}
\newcommand{\HH}{{\mathscr H}}

\newcommand{\LL}{{\mathcal L}}

\newcommand{\QQ}{{\mathcal Q}}
\newcommand{\Aa}{{\cal A}}
\newcommand{\Ba}{{\cal B}}
\newcommand{\Fa}{{\cal F}}

\newcommand{\xb}{{\boldsymbol{x}}}

\newcommand{\Mfr}{{\mathfrak M}}

\newcommand{\supp}{{\rm supp}\,}

\newcommand{\ip}[2]{{\langle #1\mid #2\rangle}}

\newcommand{\stack}[2]{\substack{#1 \\ #2}}

\newcommand{\WF}{{\rm WF}\,}

\newcommand{\Ob}{{\boldsymbol{0}}}

\newcommand{\Bb}{{\boldsymbol{B}}}
\newcommand{\Db}{{\boldsymbol{D}}}
\newcommand{\Mb}{{\boldsymbol{M}}}
\newcommand{\Nb}{{\boldsymbol{N}}}

\newcommand{\Bund}{{\sf Bund}}
\newcommand{\Ct}{{\sf C}}
\newcommand{\Man}{{\sf Man}}
\newcommand{\Set}{{\sf Set}}

\newcommand{\States}{{\sf States}}

\newcommand{\TAlg}{{\sf TAlg}}

\newcommand{\Top}{{\sf Top}}

\newcommand{\Af}{{\mathscr A}}
\newcommand{\Bf}{{\mathscr B}}
\newcommand{\Df}{{\mathscr D}}
\newcommand{\Ff}{{\mathscr F}}
\newcommand{\Gf}{{\mathscr G}}
\newcommand{\Rf}{{\mathscr R}}
\newcommand{\Sf}{{\mathscr S}}

\newcommand{\Vc}{{\mathcal V}}
\newcommand{\Wf}{{\mathscr W}}

\newcommand{\obj}{{\rm obj}\,}
\newcommand{\id}{{\rm id}}
\newcommand{\nto}{\stackrel{.}{\to}}
\newcommand{\op}{{\rm op}}
\newcommand{\Nat}{{\rm Nat}}

\newcommand{\Sp}{{\rm Sp}}

\newcommand{\cl}{{\rm cl}}
\newcommand{\co}{{\rm co}}

\begin{document}


\title{Quantum energy inequalities and local covariance II: Categorical formulation}
\author{Christopher J. Fewster\thanks{Electronic address: \tt cjf3@york.ac.uk}\\
Department of Mathematics, University of York\\ 
Heslington, York YO10 5DD, United Kingdom.}
\date{30 July 2007}

\maketitle

\begin{abstract}
We formulate Quantum Energy Inequalities (QEIs) in the framework of locally
covariant quantum field theory developed by Brunetti, Fredenhagen and
Verch, which is based on notions taken from category theory. 
This leads to a new viewpoint on the QEIs, and also to the identification
of a new structural property of locally
covariant quantum field theory, which we call Local Physical
Equivalence. Covariant formulations of the numerical range and
spectrum of locally covariant fields are given and investigated, and a
new algebra of fields is identified, in which fields are treated
independently of their realisation on particular spacetimes and 
manifestly covariant versions of the functional calculus may be
formulated.  
\end{abstract}

\section{Introduction}

A very elegant formulation of local covariance for quantum field
theory in curved spacetimes has been proposed
recently by Brunetti, Fredenhagen and Verch~\cite{BrFrVe03} [hereafter abbreviated as
BFV], utilising techniques from category
theory. Ideas of this type have already played an important role in the
proof of a rigorous spin--statistics connection in curved spacetimes~\cite{Verch01},
the perturbative renormalisation of interacting scalar field theories
in curved spacetime~\cite{Ho&Wa01,Ho&Wa02}, and the theory of superselection
sectors~\cite{Br&Ru05} and it seems that the complex of ideas set out by
BFV will have many further applications in this area. (See
also~\cite{BrPo&Ru05} for a review.) Indeed, one
should say that {\em all} structural features of interest in QFT in
CST should be formulated in this framework; any which are not capable
of such reformulation must be either discarded as noncovariant, or
(less likely, perhaps) prompt a review of the status of
covariance itself. 

This paper continues a discussion of the locally covariant aspects
of quantum energy inequalities (QEIs) that was initiated
in~\cite{Few&Pfen06}. QEIs are the remnants in QFT of the classical
energy conditions of general relativity (see~\cite{Fews05,Roma04,Few:Bros} for
recent reviews) and usually take the form of state-independent lower bounds on
suitable averages of the stress-energy tensor. In~\cite{Few&Pfen06}, it
was shown that known examples of QEIs can be formulated in a covariant fashion, and
that this could be used to obtain a priori bounds on ground state energy
densities in the Casimir effect and similar situations (see also~\cite{Marecki06}).
The presentation in~\cite{Few&Pfen06}, while influenced by BFV, did not make use of the
categorical formulation of local covariance, and it is the initial task
of this paper to show how the gap may be bridged. The aim is to isolate
the structures which might be characteristic of QEIs in general quantum field
theories in curved spacetimes, with two ends in mind: first, as a
preparatory step before attempting to derive QEIs in general covariant
quantum field theories; second, so as to provide a framework for
studying quantum field theories which are assumed to obey such bounds. 

We begin with a review of the BFV framework in
Sect.~\ref{sect:cats_for_lcQFT}, and then reexamine the definitions
of locally covariant quantum energy inequalities given in~\cite{Few&Pfen06} in this light. 
Guided by the mathematical framework, we are led to a more general
viewpoint on such bounds, and our revised definitions encompass quantum
fields other than the stress energy tensor, and permit state-dependent lower
bounds. In fact, it turns out that state-dependent lower bounds are
necessary to obtain QEIs on the non-minimally coupled scalar
field~\cite{FO07} (see also Appendix~\ref{appx:nonminimal}),
and so the latter generalisation is not merely a mathematical
extravagance. However, the new freedom would also permit rather trivial
bounds, and we therefore give a first attempt at a definition of what a nontrivial
quantum inequality should be. Inequalities on fields other than the
stress-energy tensor are also of interest; see~\cite{Marecki02} for an
application to squeezed states in quantum optics. We call these bounds
quantum inequalities (QIs). 

Investigation of our definitions quickly
reveals the desirability of a new property of locally
covariant quantum field theories, which we call {\em local physical
equivalence}. This property, described in Sect.~\ref{sect:LPE}, 
ensures that no observer can tell, by finitely many local measurements
made to finite tolerance, that the
spacetime (s)he believes (s)he inhabits is not, in fact, part of a
larger spacetime. This brings a more definite form to ideas expressed
some time ago in Kay's work on the Casimir effect~\cite{Kay79} and a
remark of Fulling in the appendix to~\cite{Fulling73}, 
which have also played a role in the development of locally covariant
QFT. In the situation at hand, local physical equivalence is needed to establish
the covariance of various constructions relating to the QIs. We also
show how information about a class of spacetimes with toroidal spatial
sections can be used to make deductions about QIs in Minkowski space.

The discussion of quantum inequalities leads naturally to a
broader consideration of the numerical
range and spectrum of a locally covariant field, in
Sect.~\ref{sec:order}. We study these objects partly for their own sake,
but also because they suggest the utility of a new algebra of fields, abstracted
from particular spacetimes or smearings. Algebras of this type offer 
manifestly covariant versions of constructions such as functional
calculus and perhaps should be considered as the natural arena for the
structural analysis of quantum fields in curved spacetime. 

Although the present contribution is largely conceptual in scope, 
the ideas which led up to it have found concrete applications in
providing the a priori bounds already mentioned~\cite{Few&Pfen06,Marecki06}, and in proving the
averaged null energy condition for the free scalar field along null geodesics
with suitable Minkowskian neighbourhoods~\cite{FewOlumPfen}. Moreover,
the local physical equivalence property and the new abstract field
algebras are of independent interest. While we do {\em not} seek to
prove new QEIs in this paper, we show in Appendix~\ref{appx:Wick_square} that
the Wick square of the free scalar field obeys 
a locally covariant difference quantum inequality (by similar arguments
to those expressed in~\cite{Few&Pfen06}) and also establish a new
result: namely that this difference quantum inequality is closely
associated with a covariant absolute quantum inequality. A number of ideas for
further study are summarised in the conclusion. Finally,
Appendix~\ref{appx:nonminimal} indicates the type of state-dependence
that occurs in a QEI for the non-minimally coupled scalar
field~\cite{FO07}.

\section{Categorical framework of locally covariant QFT}\label{sect:cats_for_lcQFT}

\subsection{Categories, functors and natural transformations}

To start, we briefly recall the definitions of some fundamental concepts from category
theory, using, for the most part, the notation and terminology
of~\cite{MacLane}. First, a {\em category} $\Ct$ consists of a set of objects
$\obj\Ct$, and, for every pair of objects $A,B$ in $\Ct$, a set $\hom_\Ct(A,B)$
of morphisms between $A$ and $B$. A morphism in $\hom_\Ct(A,B)$ is represented
diagrammatically by $A\stackrel{f}{\to} B$ or $f:A\to B$. Every object
$A\in\obj\Ct$ has a unique identity morphism $\id_A\in\hom_\Ct(A,A)$;
moreover, if $f:A\to B$ and $g:B\to C$ there is a composite morphism
$g\circ f:A\to C$ obeying the unit law
\begin{equation*}
\id_B \circ f= f\qquad g\circ\id_B = g
\end{equation*}
and associativity, $(f\circ g)\circ h=f\circ(g\circ h)$. The category
$\Set$ of small sets, with functions as morphisms, provides a standard
example. Mention of `small sets' is necessary here
because the collection of all sets is not a set, and therefore is too
large to form a category according to our definition. Following
Mac~Lane~\cite{MacLane,MacLane69} we address foundational issues by assuming the existence of a single
universe in addition to the ZFC axioms of set theory. The elements of
the universe are called {\em small sets} and serve as the objects of ordinary
mathematics, while subsets of the universe which are not also elements
of it are referred to as {\em large sets}. The advantage is that even
the large sets, and the universe itself, are sets within a model of ZFC
set theory and so one is able to manage to a large extent without ever
invoking proper classes or larger structures. Typically, the object sets
of categories we study will be large sets [for example, $\obj\Set$ is
the universe] while the sets of homomorphisms between objects will be small.

Most of the categories we study will be {\em concrete}; that
is, the objects are small sets (possibly with additional structure) and the
morphisms are functions between them.

Turning to the second key concept, a {\em
functor} $\Ff$ between categories $\Ct$ and $\Ct'$, written
$\Ff:\Ct\to \Ct'$, is a map assigning to each object $A\in\obj\Ct$ an object
$\Ff(A)\in\Ct'$ and to each morphism $f\in\hom_\Ct(A,B)$ a morphism
$\Ff(f)\in\hom_\Ct(\Ff(A),\Ff(B))$ such that 
\begin{equation*}
\Ff(\id_A) =\id_{\Ff(A)},\qquad \Ff(f'\circ f) =\Ff(f')\circ\Ff(f)
\end{equation*}
for all $A\in\obj\Ct$ and all composable morphisms $f$ and $f'$.
Functors are often described as covariant, in contradistinction to the
notion of a {\em contravariant functor} $\Ff:\Ct\to\Ct'$, which assigns objects in $\Ct'$
to objects in $\Ct$ as before, but with the assignment of morphisms running
in the opposite direction: to each $f\in\hom_\Ct(A,B)$ the functor
assigns a morphism $\Ff(f)\in\hom_\Ct(\Ff(B),\Ff(A))$, subject to the
contravariance properties 
\begin{equation*}
\Ff(\id_A) =\id_{\Ff(A)},\qquad \Ff(f'\circ f) =\Ff(f)\circ\Ff(f')\,.
\end{equation*} 
Contravariant and covariant functors are related as follows. To each category
$\Ct$ there is an {\em opposite category}, $\Ct^\op$, with the same objects as $\Ct$, 
but with all arrows reversed. That is, to each $f:A\to B$ in
$\Ct$ there is a unique $f^\op:B\to A$ in $\Ct^\op$, and every morphism
in $\Ct^\op$ arises in this way.\footnote{Clearly $f^\op:B\to A$ cannot
necessarily be identified with a {\em function} from the underlying set
of $B$ to that of $A$, so $\Ct^\op$ need not be concrete, even if $\Ct$
is.} Composition of morphisms in $\Ct^\op$ is given by the rule
$f^\op\circ g^\op = (g\circ f)^\op$ when the composition $g\circ f$ is
defined in $\Ct$. Clearly $(\Ct^\op)^\op=\Ct$. 
The assignments $A\mapsto A$, $f\mapsto f^\op$
constitute a contravariant functor $\op_\Ct:\Ct\to\Ct^\op$; any covariant
(resp., contravariant) functor $\Ff:\Ct\to\Ct'$ induces a contravariant
(resp., covariant) functor $\Ff^\op =\op_{\Ct'}\circ\Ff:\Ct\to\Ct'^\op$,
and $(\Ff^\op)^\op =\Ff$. We will make use of the freedom to replace 
contravariant functors by (covariant) functors later on. Note that,
unless otherwise specified, functors are covariant. 

We will also make use of the notion of a subfunctor. If $\Ff$ and $\Gf$
are covariant (resp., contravariant) functors from $\Ct$ to $\Ct'$,
where $\Ct'$ is a concrete category, then $\Ff$ is said to be a {\em
subfunctor} of $\Gf$ if (i) for each object $A$ in $\Ct$, the object
$\Ff(A)$ is a subset of $\Gf(A)$ and (ii) for each morphism $f:A\to B$
of $\Ct$, the morphism $\Ff(f)$ is the restriction of $\Gf(f)$ to
$\Ff(A)$, denoted $\Ff(f)=\Gf(f)|_{\Ff(A)}$ (resp., the restriction to $\Ff(B)$, 
denoted $\Ff(f)=\Gf(f)|_{\Ff(B)}$).\footnote{More precisely, $\Ct'$ is
concrete if there is a faithful functor (called the forgetful functor) from $\Ct'$ to 
$\Set$, mapping any object
of $\Ct'$ to its underlying set, and any morphism to the underlying
function. Here, a functor is faithful if its action on morphisms is injective. In
defining subfunctors, one should strictly say that $U(\Ff(A))\subset
U(\Gf(A))$ and
$U(\Ff(f))$ is the restriction of $U(\Gf(f))$, where $U$ is the
forgetful functor on $\Ct'$.} In this case, we
write $\Ff\subset\Gf$.\footnote{As with set inclusions, the use of the
symbol $\subset$ is not intended to exclude equality.} 

The third concept is the idea of a natural transformation. Suppose
$\Ff$ and $\Gf$ are functors between $\Ct$ and $\Ct'$. A {\em
natural transformation} between $\Ff$ and $\Gf$, written
$\tau:\Ff\nto\Gf$, assigns to each object $A$ of $\Ct$ a morphism
$\tau_A:\Ff(A)\to\Gf(A)$ in $\Ct'$, such that the rectangular part of
the diagram
\begin{equation*}
\begin{CD}
A @.\phantom{AAAA}@. \Ff(A) @>\tau_A>> \Gf(A)\\
@V{f}VV @.  @V{\Ff(f)}VV     @VV{\Gf(f)}V\\
B @.\phantom{AAAA}@. \Ff(B)@>>\tau_B> \Gf(B)
\end{CD}
\end{equation*}
commutes whenever $f:A\to B$ in $\Ct$, i.e., $\tau_B\circ\Ff(f)$ and
$\Gf(f)\circ\tau_A$ are identical morphisms in
$\hom_{\Ct'}(\Ff(A),\Gf(B))$.  

\subsection{Locally covariant quantum field theory}

We may now describe the structure of locally covariant quantum field
theory, largely following BFV but with some minor changes. This begins
with a category $\Man$ of spacetimes. More specifically, 
the objects of $\Man$ are $d$-dimensional, oriented and time-oriented globally
hyperbolic Lorentzian spacetimes, denoted $\Mb$, $\Nb$ etc., where the
notation denotes not just the underlying
spacetime manifold, but also the specific choices of metric and
(time-)orientation. Global hyperbolicity of $\Mb$ requires strong causality and
that $J_\Mb^+(p)\cap J^-_\Mb(q)$ is compact for all $p,q\in\Mb$~\cite{HawkingEllis}. 
The morphisms of $\Man$ are isometric embeddings
$\psi:\Mb\to\Nb$ such that (i) $\psi(\Mb)$ is an open globally hyperbolic
subset\footnote{Since $\Nb$ is globally hyperbolic, this amounts to the
requirement that, for all $p,q\in\psi(\Mb)$, each $J^+_\Nb(p)\cap
J^-_\Nb(q)$ is contained in $\psi(\Mb)$; an equivalent
formulation (given in BFV) is that every causal curve in $\Nb$ whose
endpoints lie in $\psi(\Mb)$ should be contained entirely in $\psi(\Mb)$.}
of $\Nb$, and (ii) the (time)-orientation of $\Mb$ coincides with
that pulled back from $\Nb$ via $\psi$. 

A {\em locally covariant quantum field theory} is defined to be any functor
$\Af:\Man\to\TAlg$, the category of unital topological
$*$-algebras with continuous unit-preserving faithful $*$-homomorphisms as
morphisms. That is, $\Af$ assigns to each $\Mb\in\obj\Man$ an
algebra $\Af(\Mb)$, and to each morphism $\psi\in\hom_\Man(\Mb,\Nb)$ a
faithful $*$-homomorphism $\alpha_\psi=\Af(\psi)\in\hom_{\TAlg}(\Af(\Mb),\Af(\Nb))$, so that the covariance properties
\begin{equation*}
\alpha_{\id_\Mb} =\id_{\Af(\Mb)}\qquad
\alpha_{\psi'\circ\psi}=\alpha_{\psi'}\circ\alpha_\psi
\end{equation*}
are obeyed, the latter holding whenever $\psi'$ and $\psi$ can be
composed. We will follow BFV in writing $\alpha_\psi$ for $\Af(\psi)$ to
unburden the notation. Many variations are possible, for example,
$\TAlg$ could be replaced by the category of unital $C^*$-algebras, which is the
main example in BFV.

The next step is the definition of a quantum field. We take a slightly
more general viewpoint than BFV here, considering vector-valued fields
which permit distributional smearings, instead of scalar fields smeared
with smooth compactly supported test functions. This is necessary
because QEIs are constraints on a rank-$2$ tensor field (the
stress-energy tensor) in which distributional smearings are used to
express averages along timelike submanifolds of the spacetime. 
Beginning with some
notation, if $\Bb\stackrel{\pi}{\to}\Mb$ is a smooth vector bundle (with
finite-dimensional fibre for simplicity)
then the dual bundle (whose fibres are dual to those of $\Bb$) 
is denoted $\Bb^*\stackrel{\pi^*}{\to}\Mb$; moreover, $\EE'(\Bb)$ will
denote the space of compactly supported `distributional sections' of $\Bb$, i.e., the
topological dual of $C^\infty(\Bb^*)$, the space of smooth sections of
$\Bb^*$. The smooth vector bundles $\Bb\stackrel{\pi}{\to}\Mb$ over manifolds in $\Man$ provide the
objects of a category $\Bund$, in which the morphisms between
$(\Bb_1,\pi_1,\Mb_1)$ and $(\Bb_2,\pi_2,\Mb_2)$ are pairs 
$(\zeta,\zeta^\circ)$ of smooth maps $\zeta:\Bb_1\to\Bb_2$ and
$\zeta^\circ\in\hom_{\Man}(\Mb_1,\Mb_2)$ such that $\pi_2\circ\zeta =
\zeta^\circ \circ\pi_1$ and $\zeta|_{\pi_1^{-1}(x)}:\pi_1^{-1}(x)\to\pi_2^{-1}(\zeta^\circ(x))$
is a linear isomorphism for each $x\in \Mb_1$. We say that this bundle map
{\em covers} the morphism $\zeta^\circ$. Note that our insistence
that each $\zeta|_{\pi_1^{-1}(x)}$ is a linear
isomorphism guarantees that $\Bb_1$ and $\Bb_2$ have a common fibre.
A bundle morphism $(\zeta,\zeta^\circ)\in\hom_{\Bund}(\Bb_1,\Bb_2)$ induces a push-forward
$\zeta_*$ mapping $\EE'(\Bb_1)$ to $\EE'(\Bb_2)$ 
defined by 
\begin{equation*}
(\zeta_* f)(u) = f(\zeta^* u)\,,
\end{equation*}
in terms of the pull-back of smooth sections
$(\zeta^*u)(p)=\zeta|_{\pi_1^{-1}(p)}^*(u(\zeta^\circ(p)))$. The maps
$\Bb_1\to\EE'(\Bb_1)$, $(\zeta,\zeta^\circ)\to\zeta_*$ constitute a 
functor $\EE':\Bund\to\Set$.\footnote{In many circumstances it would be
more natural to think of $\EE'$ as a functor to the category of
topological vector spaces; however we wish to consider general subsets
of testing functions in what follows, rather than subspaces.}

To specify a quantum field, the first step is to give a 
functor $\Bf:\Man\to\Bund$ satisfying the requirement that, if
$\psi:\Mb\to\Nb$, then $\Bf(\psi)$ should cover $\psi$. This functor
determines the tensor or spinor type of
the test fields. [These bundles might be associated bundles to a
spin-bundle over $\Mb$, and, strictly speaking, one should include the
choice of spin structure as part of the specification of $\Mb$ and
morphisms---see~\cite{Verch01}; we have suppressed this here.] 
A {\em covariant set of smearing fields} is any
subfunctor $\Df\subset \EE'\circ\Bf$: that is, each $\Df(\Mb)$ is a
set of compactly supported distributional sections of the bundle
$\Bf(\Mb)$, and each morphism $\psi:\Mb\to\Nb$ in $\Man$ has a
push-forward action $\Df(\psi)=\Bf(\psi)_*|_{\Df(\Mb)}$ injectively mapping
$\Df(\Mb)$ into $\Df(\Nb)$. To unburden the
notation, we write $\psi_*$ for $\Df(\psi)$. 
Each $\Df(\Mb)$ will be the class of test sections against which the
quantum field (to be defined next) will be smeared on spacetime $\Mb$. A
simple example is provided by the scalar field, where we take
$\Df(\Mb)=\CoinX{\Mb}$.\footnote{Here, the underlying bundle is
$\Bf(\Mb)=\Mb\times\CC$ and $\EE'(\Mb\times\CC)$ is identified with
the usual class of compactly supported distributions on $\Mb$, into
which $\CoinX{\Mb}$ may be embedded using the metric-induced volume form on $\Mb$.} On the other
hand, Dimock's quantisation of the electromagnetic field~\cite{Dimock92} (see
also~\cite{Few&Pfen03}) provides an example
where the set of smearing fields is restricted, in that case to the
divergence-free one-forms on $\Mb$. 

A {\em locally covariant quantum field} can now be described as a natural transformation
$\Phi:\Df\nto\Af$ between a locally covariant set of smearing fields and a locally
covariant quantum field theory, represented by functors $\Df$ and $\Af$
as above, with $\TAlg$ regarded as
a subcategory of $\Set$.\footnote{More precisely, $\Phi$ is a natural
transformation between $\Df$ and $U\Af$, where $U:\TAlg\to\Set$ is the
forgetful functor.} Namely, to each $\Mb$ we associate a (not
necessarily linear or continuous) map $\Phi_\Mb:\Df(\Mb)\to\Af(\Mb)$ so that the
rectangle in
\begin{equation*}
\begin{CD}
\Mb @.\phantom{AAAA}@.\Df(\Mb) @>\Phi_{\Mb}>> \Af(\Mb)\\
@V{\psi}VV @. @V{\psi_*}VV     @VV{\alpha_{\psi}}V\\
\Nb @.\phantom{AAAA}@.\Df(\Nb)@>>\Phi_{\Nb}> \Af(\Nb)
\end{CD}
\end{equation*}
commutes, i.e., $\alpha_\psi(\Phi_\Mb(f)) = \Phi_\Nb(\psi_*f)$, where
(as above) we have
written $\psi_*$ for the bundle morphism $\Df(\psi)$.

This definition is a slight generalisation of that proposed by BFV, in
that BFV only considered scalar fields (but see~\cite{Verch01})
in the case where $\Df(\Mb)$ is the space of smooth compactly supported
functions on $\Mb$, rather than a 
subset of the space of compactly supported distributional sections of a bundle.
We have also
formulated the natural transformation within $\Set$, rather than the
category $\Top$ of topological spaces; this amounts to dropping the
continuity condition on $\Phi_\Mb$. On the other hand, BFV explicitly
envisaged the possibility that $\Phi_\Mb$ might not be linear, so as to
accommodate objects such as local $S$-matrices. 

The final piece of general structure we will need is the concept of a
locally covariant state space. If $\Aa$ is a unital topological $*$-algebra, its state
space, denoted $\Aa^*_{+,1}$ is the convex set of positive
($\omega(A^*A)\ge 0$), normalised ($\omega(\II)=1$) continuous linear
functionals $\omega:\Aa\to\CC$. We endow $\Aa^*_{+,1}$ with the weak-$*$
topology. The set of all
states is generally too large for physical purposes, so it is convenient
to refer to any convex subset $S\subset\Aa^*_{+,1}$, with the subspace
topology, as a state space for $\Aa$. Such subsets will form the objects
of a category $\States$. The morphisms in $\States$ will be all
continuous affine maps, i.e.,
$\LL\in\hom_\States(S,S')$ if $\LL:S\to S'$ is continuous and obeys
$\LL(\lambda\omega+(1-\lambda)\omega')=\lambda\LL\omega+(1-\lambda)\LL\omega'$
for all $\omega,\omega'\in S$, $\lambda\in[0,1]$. Our definitions
differ slightly from those in BFV, who required that $S$ should
be closed under operations induced by $\Aa$ and only considered
morphisms which arise as duals of $*$-algebra monomorphisms. The latter requirement will
enter in our definition of the state space functor, so there is no
essential difference in the present discussion.

Naturally associated with any functor $\Af:\Man\to\TAlg$, there is a
contravariant functor $\Af^*_{+,1}:\Man\to\States$ given by
$\Af^*_{+,1}(\Mb)=\Af(\Mb)^*_{+,1}$ and
$\Af^*_{+,1}(\psi)=\alpha_\psi^*|_{\Af^*_{+,1}(\Mb)}$. We define a
locally covariant {\em state space} for the theory $\Af$ to be any
(contravariant) subfunctor $\Sf\subset\Af^*_{+,1}$. Thus each 
$\Sf(\Mb)$ is a convex subset of states
on the algebra $\Af(\Mb)$ assigned to $\Mb$, and, for any
$\psi:\Mb\to\Nb$ we have $\Sf(\psi)=\alpha_\psi^*|_{\Sf(\Nb)}$, where
$\alpha_\psi=\Af(\psi)$ is the faithful $*$-homomorphism between
$\Af(\Mb)$ and $\Af(\Nb)$ induced by $\psi$. 

The various structures introduced so far interact in the following way.
Let $\Phi$ be a locally covariant quantum field associated with the
locally covariant QFT $\Af$ and smearing fields $\Df$. Suppose
$\psi:\Mb\to\Nb$ in $\Man$ and that $\omega\in\Sf(\Nb)$. Then there is 
a state $\alpha_\psi^*\omega\in\Sf(\Mb)$ with $n$-point
function\footnote{We use the term slightly loosely in the situation
where $\Phi_\Mb$ is not linear.}
\begin{equation*}
\alpha_\psi^*\omega(\Phi_\Mb(f_1)\cdots\Phi_\Mb(f_n)) = 
\omega(\alpha_\psi(\Phi_\Mb(f_1)\cdots\Phi_\Mb(f_n))) = 
\omega(\Phi_\Nb(\psi_*f_1)\cdots\Phi_\Nb(\psi_*f_n))\,;
\end{equation*}
that is, the $n$-point function of $\alpha_\psi^*\omega$ on $\Mb$ is the
pull-back of the $n$-point function of $\omega$ on $\Nb$.

A key example to bear in mind is that of the Hadamard states of the
free scalar field, which are distinguished by the wave-front set of the
two-point function. Since the wave-front set transforms in a natural
fashion under the pull-back of distributions, we indeed have the
embedding $\alpha_\psi^*\Sf(\Nb)\subset\Sf(\Mb)$.

\section{Locally covariant quantum inequalities}\label{sect:lcQIs}

\subsection{Absolute quantum inequalities}\label{sect:AQIs}

The stress-energy tensor of classical matter is usually taken to obey
certain {\em energy conditions}. For example, $T_{ab}$ obeys the {\em weak
energy condition} if $T_{ab}u^au^b\ge 0$ for all timelike $u^a$, which
means that all observers detect nonnegative energy density. It is not
possible for such conditions to hold in quantum field theory 
at individual points~\cite{EGJ}. In some models, however, local averages of the
expectation value of the
stress-energy tensor can be bounded from below as functions of the
state, and these bounds constitute Quantum Energy Inequalities (QEIs).
Inequalities of this type have been obtained for free fields in flat and
curved spacetimes (see, e.g., \cite{Fews05,Roma04,Few:Bros} for
discussion and references, and \cite{FewSmith07,Smith07} for recent results)
and for two-dimensional conformal field theories in
Minkowski space~\cite{Fe&Ho05}. (A second type of QEI, to be discussed
in the next section, has been proved in many more cases.) 
In~\cite{Few&Pfen06} a notion of a locally covariant
QEI was formulated without fully locating it within the categorical structures
introduced by BFV; our purpose in this section is to remedy this and
to explore some generalisations suggested in the process. 

Suppose, then, that a quantum field theory, represented by a functor
$\Af:\Man\to\TAlg$ has a stress-energy tensor $T$. That is, there should
be a functor $\Df:\Man\to\Set$, with $\CoinX{T_0^2\Mb}\subset\Df(\Mb)\subset
\EE'(T_0^2\Mb)$, i.e., compactly supported distributional
contravariant tensor fields of rank two, so that $T:\Df\nto\Af$. (The field $T$ should also
satisfy conditions which identify it as the stress-energy tensor of the
theory --- see, e.g., the discussion in BFV --- but we will not need these conditions here.)
In~\cite{Few&Pfen06}, a locally covariant QEI was defined to be an
assignment, to each $\Mb\in\Man$, of a subset
$\Ff(\Mb)\subset\EE'(T_0^2\Mb)$ and a function
$\widetilde{\QQ}_\Mb:\Ff(\Mb)\to\RR$ such that
\begin{equation}
\omega(T_\Mb(f))\ge -\widetilde{\QQ}_\Mb(f)
\label{eq:protoQEI}
\end{equation}
for all $\omega\in\Sf(\Mb)$; moreover, if $\psi:\Mb\to\Nb$ in $\Man$
then $\psi_*\Ff(\Mb)\subset\Ff(\Nb)$ and 
\begin{equation}
\widetilde{\QQ}_\Nb(\psi_*f) = \widetilde{\QQ}_\Mb(f)
\label{eq:Qt_trans}
\end{equation}
for all $f\in\Ff(\Mb)$. Clearly, one would also require
$\Ff(\Mb)\subset\Df(\Mb)$ if the latter is a proper subset of
$\EE'(T_0^2\Mb)$. More precisely, this was the definition of an {\em absolute}
QEI, by contrast with the {\em difference} QEIs to be discussed later. 
Note that one needs the freedom to restrict the class of smearings
$\Ff(\Mb)$ because, even classically, not all smearings of the
stress-energy tensor are expected to be semi-bounded. 

Comparing this definition with the general structures described above,
it is clear that the assignment $\Mb\mapsto\Ff(\Mb)$,
$\psi\mapsto\Ff(\psi)=\psi_*$ defines a functor
$\Ff:\Man\to\Set$ which is a subfunctor of $\Df$.
In addition, Eq.~\eqref{eq:Qt_trans} strongly suggests that each
$\widetilde{\QQ}_\Mb$ is a component of a natural transformation between
$\Ff$ and some other functor from $\Man$ to (a subcategory of) $\Set$.
One way of making this precise is to define a constant functor $\Rf:\Man\to\Set$ by $\Rf(\Mb)=\RR$
for all objects $\Mb\in\Man$ and 
$\Rf(\psi)=\id_{\RR}$ for all morphisms $\psi$ of $\Mb$. Then the
rectangular portion of
\begin{equation*}
\begin{CD}
\Mb @.\phantom{AAAA}@.\Ff(\Mb) @>\widetilde{\QQ}_{\Mb}>> \RR=\Rf(\Mb) \\
@V{\psi}VV @. @V{\psi_*}VV     @VV{\id_{\RR}}V\\
\Nb @.\phantom{AAAA}@.\Ff(\Nb)@>>\widetilde{\QQ}_{\Nb}> \RR=\Rf(\Nb)
\end{CD}
\end{equation*}
commutes, so the $\widetilde{\QQ}_\Mb$ are indeed the components of a
natural transformation $\widetilde{\QQ}:\Ff\nto\Rf$. 

However, it 
is known that there are quantum field theory models that do not obey
QEIs of the above type. An example is given by the non-minimally coupled
scalar field, for which it may be proved that the renormalised energy
density is unbounded from below even after averaging \cite{FO07}.
The same is also expected for general interacting theories for reasons
described in \cite{OlumGraham03}. For these models, therefore, no
$\widetilde{\QQ}_\Mb$ of the required type exists. We therefore broaden the notion
of an absolute QEI to permit state-depdendent lower bounds of the form
\begin{equation*}
\omega(T_\Mb(f))\ge -\omega(\QQ_\Mb(f)) \quad\text{for all
$\omega\in\Sf(\Mb)$,}
\end{equation*}
where $\QQ_\Mb$ is now an element of the algebra $\Af(\Mb)$. This must
be supplemented by conditions that exclude trivial bounds (such as that obtained by setting
$\QQ_\Mb(f)=-T_\Mb(f)$); a first attempt at this will be described later. 
As before, covariance of the QEI is
represented by the requirement that the $\QQ_\Mb$ be components of a
natural transformation, but now the transformation is between $\Ff$ and
$\Af$, and is represented by the diagram
\begin{equation}
\begin{CD}
\Mb @.\phantom{AAAA}@.\Ff(\Mb) @>\QQ_{\Mb}>> \Af(\Mb) \\
@V{\psi}VV @. @V{\psi_*}VV     @VV{\alpha_\psi}V\\
\Nb @.\phantom{AAAA}@.\Ff(\Nb)@>>\QQ_{\Nb}> \Af(\Nb).
\end{CD}
\label{eq:QInatural}
\end{equation}
Note that a state-independent QEI of the form~\eqref{eq:protoQEI}
is accommodated within our new definition 
by setting
\begin{equation*}
\QQ_\Mb(f) = \widetilde{\QQ}_\Mb(f)\II_{\Af(\Mb)}\,;
\end{equation*}
because $\omega(\QQ_\Mb(f))=\widetilde{\QQ}_\Mb(f)$ for any state $\omega$ and, 
recalling that any $\alpha_\psi$ is unit-preserving, we have
\begin{equation*}
\alpha_\psi(\QQ_\Mb(f)) =
\widetilde{\QQ}_\Mb(f)\alpha_\psi(\II_{\Af(\Mb)})=
\widetilde{\QQ}_\Nb(\psi_*f) \II_{\Af(\Nb)}=\QQ_\Nb(\psi_*f)
\end{equation*}
whenever $\psi:\Mb\to\Nb$ in $\Man$.

We may now give our definition of a locally covariant absolute quantum
inequality, which incorporates the above discussion and also allows for
the possibility that fields other than the stress-energy tensor may be
subject to similar bounds. 

\begin{Def} Let $\Phi$ be a locally covariant quantum field associated
with a locally covariant QFT $\Af$ and covariant set of smearing fields $\Df$. A {\em locally
covariant absolute quantum inequality} on $\Phi$ relative to a state space
$\Sf$ consists of a subfunctor $\Ff\subset \Df$ and a natural
transformation $\QQ:\Ff\nto\Af$ such that
\begin{equation*}
\omega(\Phi_\Mb(f)+\QQ_\Mb(f)) \ge 0
\end{equation*}
 for all $\omega\in\Sf(\Mb)$ and $f\in\Ff(\Mb)$.
\end{Def}
The naturality condition requires, of course, that
\begin{equation*}
\alpha_\psi(\QQ_\Mb(f)) = \QQ_\Nb(\psi_*f)
\end{equation*}
for all $f\in\Ff(\Mb)$, whenever $\psi:\Mb\to\Nb$ in $\Man$. We will
sometimes use AQI for `absolute quantum inequality'. 

Two remarks are in order. First, 
the continuity properties of the map $f\mapsto\QQ_\Mb(f)$ are not
fully understood in known examples of QEIs; 
this is why we chose to formulate the notion of a
locally covariant field [and hence locally covariant QIs] 
without continuity assumptions, that is, in $\Set$, rather than $\Top$. In due course
one might hope for a more finely grained definition.

Second, as already mentioned, we do not assume that the
elements $\QQ_\Mb(f)\in\Af(\Mb)$ are scalar multiples of the identity,
in contrast to the usual QEI literature. The situation where $\QQ_\Mb(f)$
is proportional to the identity is evidently very attractive, 
because it produces state-independent lower bounds. However, our more
flexible definition has the advantage of encompassing theories such as
the non-minimally coupled scalar field, which do not obey state-independent
QEIs, but do obey more general bounds~\cite{FO07} as we will briefly
discuss in Appendix~\ref{appx:nonminimal}. In addition, our definition has a natural expression
in terms of an order relation among the fields of the theory -- see
Sect.~\ref{sec:order}. Having said this, we should clearly place
some restrictions on $\QQ$: for example, taking $\Ff=\Df$ and
$\QQ_\Mb(f)=-\Phi_\Mb(f)$ rather trivially satisfies the definition. 
More generally, a quantum inequality of the above type will be called
trivial on $\Mb$ if, for each $f\in\Ff(\Mb)$, there are constants $C_{\Mb,f}$ and $C_{\Mb,f}'$
(with possibly different engineering dimensions) such that
\begin{equation}
\left|\omega(\Phi_\Mb(f))\right| \le C_{\Mb,f}\left|\omega(\QQ_\Mb(f))\right| +
C_{\Mb,f}'
\label{eq:triviality_abs}
\end{equation}
for all $\omega\in\Sf(\Mb)$. In this case, of course, there is nothing
special about $\QQ$ as a {\em lower} bound. In section~\ref{sect:LPE} we will prove that our
definition of triviality is respected by local covariance (subject to the condition of Local
Physical Equivalence, introduced below). 

To illustrate the content of our definition of (non)triviality, note first that state-independent QIs (for
which the right-hand side of~\eqref{eq:triviality_abs} is independent of
$\omega$) are always nontrivial unless $\Phi_\Mb(f)$ has bounded
expectation values on $\Sf(\Mb)$. More generally, one might attempt to
introduce a notion of the `order' of a field $\Phi_\Mb$ as the infimum of those $\alpha\in\RR^+$
so that for each $f\in\Df(\Mb)$ there exist constants $C_{\Mb,f}$, $C_{\Mb,f}'$
such that 
\begin{equation*}
\left|\omega(\Phi_\Mb(f))\right| \le C_{\Mb,f}\left(\mathcal{W}_{\Mb,f}(\omega)\right)^\alpha +
C_{\Mb,f}'
\end{equation*}
for all $\omega$ in (some suitable subset of) $\Sf(\Mb)$, where
$\mathcal{W}_{\Mb,f}(\omega)$ is a measure of the energy
content\footnote{This might be based on local thermodynamic observables,
cf.~\cite{Buchholz_Ojima_Roos02,BuchSchlem07}.} of the support of $f$
or its immediate surroundings in state $\omega$; we assume $\mathcal{W}_{\Mb,f}(\omega)$ can become
arbitrarily large as $\omega$ varies. Fields that are scalar multiples of the
identity would have order zero according to this definition. If a field 
can be bounded from below by a field of lower order, this would
constitute a nontrivial quantum inequality according to our previous
definition. One could also sharpen the definition of nontriviality in
order to require the converse, i.e., that the lower bound must be of
lower order to count as nontrivial. However a full investigation of the
concept of `order' sketched above goes beyond the scope of this paper,
and we will work with the original definition of nontriviality here.
We will return briefly to these issues in the conclusion.

As a separate example, although not in
the covariant framework, consider an operator of the form
$T=\sum_{i=1}^\infty \lambda_i a_i^*a_i$ on the usual Fock space with
annihilation and creation operators obeying
$[a_i,a_j^*]=\delta_{ij}\II$. If the $\lambda_i$ are bounded from below,
say by $\lambda_0$, then we have a QI
\begin{equation*}
\ip{\psi}{T\psi} \ge \lambda_0\ip{\psi}{N\psi}
\end{equation*}
for all states $\psi$ that are finite linear combinations of states
created from the Fock vacuum by finitely many creation operators. Here
$N=\sum_{i=1}^\infty a_i^*a_i$ is the usual number operator. This bound
is nontrivial in the above sense if and only if the $\lambda_i$ are unbounded from
above. 

As a digression, we observe that quantum inequalities, as we have
defined them here, are strongly reminiscent of the
G{\aa}rding inequalities arising in the study of pseudodifferential
operators, in which one may obtain lower bounds `with a gain of two
derivatives'. For example, a (nonzero) pseudodifferential operator of
order $m$ with a nonnegative symbol can be bounded from below
by an operator of order $m-2$ but
cannot be bounded from above in this way~\cite{FefPho78}. Moreover, G{\aa}rding
inequalities are closely related to the uncertainty principle, and
provide a class of quantum mechanical quantum inequalities~\cite{EvFew&Ve05}.
It is tempting to speculate that this link might run more deeply, and might indeed suggest an
approach to quantum inequalities via the phase space properties of
quantum field theory, in which the `gain of derivatives' can be
explicitly linked to nontriviality of the QI. Further evidence and comments in this direction
may be found in~\cite{Few:Bros,FOP}. It also supports the contention that our definition of
quantum inequalities is natural from a mathematical viewpoint.

Finally, it is important to mention two examples of 
locally covariant absolute quantum energy inequalities. First,
Flanagan's bound~\cite{Flan02} for massless
scalar fields in two dimensions is covariant in this
sense~\cite{Few&Pfen06}; second, the scalar field in four dimensions
admits locally covariant absolute QEIs~\cite{FewSmith07} and the same is expected
for other free field theories, e.g., by combining the absolute QEI for
the Dirac field obtained in~\cite{Smith07} with the discussion of
covariance in~\cite{DawsFews06,FewSmith07}. 

\subsection{Difference quantum inequalities}\label{sect:DQIs}

Much of the literature on QEIs concentrates on so-called {\em difference}
inequalities, rather than the absolute bounds just discussed. In the
state-independent case, which has been the main focus of the literature, 
difference quantum inequalities are statements of the form 
\begin{equation}
\omega(\Phi_\Mb(f))-\omega_0(\Phi_\Mb(f))\ge -\widetilde{\QQ}_\Mb(f,\omega_0)\,,
\label{eq:diffQI_trad}
\end{equation}
required to hold for a class of sampling functions $f$ and states
$\omega$ and $\omega_0$. Here, $\omega_0$ is known as the reference
state and $\omega$ as the state of interest. Historically, these bounds proved the easiest to obtain in curved
spacetimes. As we have already allowed absolute QIs to depend on the
state of interest, we should extend the same freedom to difference QIs.
Thus, for our purposes, a difference QI will be a bound of the form
\begin{equation*}
\omega(\Phi_\Mb(f))-\omega_0(\Phi_\Mb(f))\ge -
\omega(\QQ_\Mb(f,\omega_0))\,,
\end{equation*}
holding for all $f$, $\omega$ and $\omega_0$ in appropriate classes.

To formulate difference QIs in categorical terms, we introduce an
additional concept. The product $\Ct_1\times\Ct_2$
of categories $\Ct_1$ and $\Ct_2$ has objects which are pairs $\langle
A_1,A_2\rangle$ of objects $A_i\in\obj\Ct_i$. Morphisms between $\langle
A_1,A_2\rangle$ and $\langle B_1,B_2\rangle$ are pairs $\langle
f_1,f_2\rangle$ of morphisms $f_i\in\hom_{\Ct_i}(A_i, B_i)$; composition
of morphisms being defined by $\langle f_1,f_2\rangle\circ\langle
f'_1,f'_2\rangle= \langle f_1\circ f_1',f_2\circ f_2'\rangle$. Moreover,
given functors $\Ff_i:\Ct\to\Ct_i$, we obtain a functor
$\langle\Ff_1,\Ff_2\rangle:\Ct\to \Ct_1\times\Ct_2$ by 
\begin{equation*}
\langle\Ff_1,\Ff_2\rangle(A) = \langle
\Ff_1(A),\Ff_2(A)\rangle \qquad
\langle\Ff_1,\Ff_2\rangle(f) = \langle
\Ff_1(f),\Ff_2(f)\rangle
\end{equation*}
on objects and morphisms respectively. 

As a particular example, in a locally covariant quantum field theory
$\Phi:\DD\nto\Af$ with state space $\Sf$, the functor
$\langle\DD,\Sf^\op\rangle:\Man\to \Set\times\States^\op$ maps each manifold
$\Mb$ to the pair $\langle \DD(\Mb),\Sf(\Mb)\rangle$ and any morphism
$\psi:\Mb\to\Nb$ to the pair of morphisms $\langle
\psi_*,\alpha_\psi^{*\op}\rangle$. We may now give our formal definition.

\begin{Def} Let $\Phi$ be a locally covariant quantum field associated
with a locally covariant QFT $\Af$ and covariant set of smearing fields $\Df$. A {\em locally
covariant difference quantum inequality} on $\Phi$ relative to a state space
$\Sf$ consists of a subfunctor $\Ff\subset \Df$ and a natural
transformation $\QQ:\langle\Ff,\Sf^\op\rangle\nto\Af$ such that
\begin{equation}
\omega(\Phi_\Mb(f))-\omega_0(\Phi_\Mb(f))\ge -
\omega(\QQ_\Mb(f,\omega_0))\,,
\label{eq:diffQI}
\end{equation}
holds for all $f\in\Ff(\Mb)$, $\omega,\omega_0\in\Sf(\Mb)$.
\end{Def}
Here, naturality requires that the rectangle in
\begin{equation*}
\begin{CD}
\Mb @.\phantom{AAAA}@.\langle\Ff(\Mb),\Sf(\Mb)\rangle @>\QQ_{\Mb}>> \Af(\Mb)\\
@V{\psi}VV @. @V{\langle\psi_*,\alpha_\psi^{*\op}\rangle}VV     @VV{\alpha_{\psi}}V\\
\Nb @.\phantom{AAAA}@.\langle\Ff(\Nb),\Sf(\Nb)\rangle @>>\QQ_{\Nb}> \Af(\Nb)
\end{CD}
\end{equation*}
commutes, that is,
\begin{equation*}
\alpha_\psi(\QQ_\Mb(f,\alpha_\psi^*\omega_0)) =
\QQ_\Nb(\psi_*f,\omega_0)\,.
\end{equation*}
We will sometimes use superscripts $d$ and $a$ to distinguish difference
and absolute QIs, and abbreviate `difference quantum inequality' as DQI. 

Many of the comments made regarding absolute QIs apply here also. In
particular, as mentioned above, one is often interested in the `state independent'
situation where $\QQ_\Mb(f,\omega_0)=\widetilde{\QQ}_\Mb(f,\omega_0)\II_{\Af(\Mb)}$,
and the naturality requirement becomes
\begin{equation*}
\widetilde{\QQ}_\Mb(f,\alpha_\psi^*\omega_0) =
\widetilde{\QQ}_\Nb(\psi_*f,\omega_0)\,,
\end{equation*}
which was the definition adopted in~\cite{Few&Pfen06} for a locally
covariant difference QI. Setting $\omega_0=\omega$ in Eq.~\eqref{eq:diffQI},
it is clear that $\widetilde{\QQ}_\Mb(f,\omega)\ge 0$ for all $f\in\Ff(\Mb)$,
$\omega\in\Sf(\Mb)$. 

It is also clear that one may have rather trivial difference QIs, such
as that obtained by setting $\QQ_\Mb(f,\omega_0) =
-\Phi_\Mb(f)+\omega_0(\Phi_\Mb(f))\II$. More generally, we define a
difference QI to be trivial on $\Mb$ if, for all $f\in\Ff(\Mb)$ and $\omega_0\in\Sf(\Mb)$,
there exist constants $C_{\Mb,f,\omega_0}$ and $C'_{\Mb,f,\omega_0}$
such that
\begin{equation}
\left|\omega(\Phi_\Mb(f))\right| \le C_{\Mb,f,\omega_0}\left|\omega(\QQ_\Mb(f,\omega_0))\right| +
C_{\Mb,f,\omega_0}'\,.
\label{eq:triviality_diff}
\end{equation}
Again, all state-independent QIs are nontrivial. 
We also emphasise that difference QEIs conforming to our definition of local
covariance are known~\cite{Few&Pfen06}. 

There is a close relationship between difference and absolute QIs. In
particular, given any AQI $\QQ^a:\Ff\nto\Af$, define
\begin{equation}
\QQ^d_\Mb(f,\omega_0) = \QQ^a_\Mb(f)
+\omega_0(\Phi_\Mb(f))\II_{\Af(\Mb)}
\label{eq:AQI_to_DQI}
\end{equation}
for all $f\in\Ff(\Mb)$, $\omega_0\in\Sf(\Mb)$ and $\Mb\in\Man$. Local
covariance of $\QQ^d$ is easily checked: if $\psi:\Mb\to\Nb$ in $\Man$
then
\begin{eqnarray*}
\QQ^d_\Nb(\psi_*f,\omega_0) &=& \QQ^a_\Nb(\psi_*f) +
\omega_0(\Phi_\Nb(\psi_*f))\II_{\Af(\Nb)}
= \alpha_\psi(\QQ^a_\Mb(f) +
\alpha_\psi^*\omega_0(\Phi_\Mb(f))\II_{\Af(\Mb)}) \\
&=&\alpha_\psi\QQ^d_\Mb(f,\alpha_\psi^*\omega_0)\,,
\end{eqnarray*}
so $\QQ^d:\langle\Ff,\Sf^\op\rangle\nto\Af$. Moreover, $\QQ^d$ is a DQI as the
elementary calculation
\begin{equation*}
\omega(\Phi_\Mb(f))-\omega_0(\Phi_\Mb(f))\ge -\omega(\QQ^a_\Mb(f))-\omega_0(\Phi_\Mb(f))
=-\omega(\QQ^d_\Mb(f,\omega_0))
\end{equation*}
shows, using the fact that $\QQ^a$ is an AQI. One may also check that
$\QQ^d$, so defined, is a nontrivial DQI on spacetime $\Mb$ if and only if $\QQ^a$
is a nontrivial AQI on $\Mb$. 

Conversely, we may start with a
locally covariant DQI $\QQ^d$, and attempt to construct an AQI. At
first sight this appears to be a simple matter of algebra: 
in any given spacetime $\Mb$, and for any state
$\omega_0\in\Sf(\Mb)$, we have
\begin{equation*}
\omega(\Phi_\Mb(f)) \ge -\omega(\QQ^d_\Mb(f,\omega_0)-\omega_0(\Phi_\Mb(f))\II_{\Af(\Mb)})
\end{equation*}
holding for all $\omega\in\Sf(\Mb)$, which suggests the simple
rearrangement of \eqref{eq:AQI_to_DQI}
\begin{equation}
\QQ^a_\Mb(f) = \QQ^d_\Mb(f,\omega_0)
-\omega_0(\Phi_\Mb(f))\II_{\Af(\Mb)}\,.
\label{eq:DQI_to_AQI}
\end{equation}
However, as noted by BFV, there is
no way of covariantly specifying a single preferred state in every
spacetime, so nontrivial dependence of the right-hand side on
$\omega_0\in\Sf(\Mb)$ would present an obstruction to local covariance
of the bound. Thus simple rearrangement allows one to pass from DQI to
an AQI if and only if 
\begin{equation}
\QQ^d_\Mb(f,\omega_0)-\QQ^d_\Mb(f,\omega_1)=
\left(\omega_0(\Phi_\Mb(f))-\omega_1(\Phi_\Mb(f))\right)\II_{\Af(\Mb)}
\label{eq:independence}
\end{equation}
for all $f\in\Ff(\Mb)$ and $\omega_0,\omega_1\in\Sf(\Mb)$. This is
quite a strong condition, and it is remarkable that it holds for one
of the main DQIs available in curved spacetimes, as we show in
Appendix~\ref{appx:Wick_square}. In general one would not have this independence, in which
case the obvious approach is to take an infimum over
$\omega_0\in\Sf(\Mb)$ in \eqref{eq:DQI_to_AQI}. Again covariance must
be checked, and this turns out to need the new ingredient of local
physical equivalence, to which we now turn.

\section{Local physical equivalence}\label{sect:LPE}

Our ability to probe physical systems with experiments is necessarily
limited to a finite number of measurements made to finite tolerance.
There is therefore good reason to regard two states as physically
equivalent if they cannot be distinguished by tests of this type (see,
for example, the lucid discussion in~\cite{Araki}). Similarly, two
state spaces may be regarded as physically equivalent if the
expectation values (of any finite set of observables) in any state in one may be
arbitrarily well-approximated by those corresponding to states in the other, and vice versa.
Technically, this is equivalent to the two state spaces having equal
closure in the weak-$*$ topology on the set of all states.

Now consider a locally covariant quantum field theory $\Af:\Man\to\TAlg$. If
$\psi:\Mb\to\Nb$, the map $\alpha_\psi^*|_{\Sf(\Nb)}$ sends each state
$\omega\in\Sf(\Nb)$ of the theory on $\Nb$ to a state
$\alpha_\psi^*\omega\in\Sf(\Mb)$ of the theory on $\Mb$. However, there is no
reason to suppose that this map is invertible, and indeed examples are
known where it is not (see, e.g., the end of section II.B in~\cite{Few&Pfen06}).
Thus there can be `more' states available to us
on the spacetime $\Mb$ than on the spacetime $\Nb$ into which it is embedded. 

However, the principle of locality should surely prevent us from
determining, by local experiments, whether we truly live in $\Mb$, or on
its image embedded in $\Nb$. Thus we should not be able to detect the `extra'
states on $\Mb$, which suggests the following requirement on the state space. 

\begin{Def} $\Sf$ respects {\em local physical equivalence} if, whenever
$\psi:\Mb\to\Nb$, $\alpha_\psi^*\Sf(\Nb)$ and $\Sf(\Mb)$ have equal
closures in the weak-$*$ topology on $\Af(\Mb)^*$. 
\end{Def}

This principle has not previously been identified in locally covariant quantum
field theory and it is therefore necessary to check whether it holds in
known models. To make a start, let us consider the situation in which 
each $\Af(\Mb)$ is a $C^*$-algebra, and each
$\Sf(\Mb)$ is closed under operations induced by $\Af(\Mb)$. That is,
for any $\omega\in\Sf(\Mb)$ and $A\in\Af(\Mb)$ for which
$\omega(A^*A)>0$, we have $\omega_A\in\Sf(\Mb)$, where
$\omega_A(B)=\omega(A^*BA)/\omega(A^*A)$. We will say that $\Sf$ is
closed under operations induced by $\Af$ in this case. This was the main focus in BFV.
In this setting we have the following:

\begin{Lem}\label{lem:Fell}
If $\Sf$ is closed under operations induced by $\Af(\Mb)$, each
$\Af(\Mb)$ is a $C^*$-algebra and each $\Sf(\Mb)$ contains at least one state inducing a faithful GNS
representation of $\Af(\Mb)$ then each $\Sf(\Mb)$ is weak-$*$ dense in the
set of all states on $\Af(\Mb)$. 
\end{Lem}
{\noindent\em Proof:} Let $\varphi\in\Sf(\Mb)$ induce a faithful
representation of $\Af(\Mb)$, and suppose $\omega$ is an arbitrary
state on $\Af(\Mb)$. By Fell's theorem (Theorem~1.2
in~\cite{Fell60}) states induced by finite rank density
matrices in the GNS representation of $\varphi$ are weak-$*$ dense in
$\Af(\Mb)^*_{+,1}$. Given any $\epsilon>0$ then, we may find nonzero
vectors $\psi_1,\ldots,\psi_N\in\HH_{\varphi}$
obeying $\sum_{r=1}^N\|\psi_r\|^2=1$ such that 
\begin{equation*}
\left|\omega(A)-\sum_{r=1}^N \ip{\psi_r}{\pi_{\varphi}(A)\psi_r}\right|<\epsilon;
\end{equation*}
furthermore, because the GNS representation is cyclic, the $\psi_r$ may
be chosen, without loss of generality, to take the form $\psi_r=\pi_{\varphi}(A_r)\Omega_{\varphi}$
for $A_1,\ldots,A_N\in\Af(\Mb)$. The sum in the last equation can be
reexpressed as $\omega'(A)$, where $\omega'=\sum_{r=1}^N \varphi(A_r^*A_r)\varphi_{A_r}$.
As $\sum_{r=1}^N \varphi(A_r^*A_r)=\sum_{r=1}^N \|\psi_r\|^2 = 1$,
$\omega'$ is a finite convex combination of states obtained from
$\varphi\in\Sf(\Mb)$ by operations in $\Af(\Mb)$, and therefore
belongs to $\Sf(\Mb)$. Accordingly, $\omega$
is a weak-$*$ limit of states in $\Sf(\Mb)$. $\square$

\begin{Prop} \label{prop:LPE_from_Fell}
Consider a locally covariant quantum field theory in which each $\Af(\Mb)$ is a
$C^*$-algebra and $\Sf$ is closed under operations induced by $\Af$. If each
$\Sf(\Mb)$ contains at least one state inducing a faithful GNS
representation of $\Af(\Mb)$ then $\Sf$ respects local physical
equivalence. In particular, if each $\Af(\Mb)$ is simple and each $\Sf(\Mb)$
is nonempty then $\Sf$ respects local physical equivalence.
\end{Prop}
{\noindent\em Proof:} It is sufficient to show that, for every morphism
$\psi:\Mb\to\Nb$ in $\Man$, $\Sf(\Mb)$ is contained in the weak-$*$ closure of
$\alpha_\psi^*\Sf(\Nb)$ (the reverse inclusion is trivial). 
Accordingly, let $\omega$ be any state in $\Sf(\Mb)$. 
As $\alpha_\psi$ is a $*$-morphism of
$C^*$-algebras, $\alpha_\psi(\Af(\Mb))$ is a $C^*$-subalgebra of
$\Af(\Nb)$ (Prop.~2.3.1 in~\cite{BratRob}) on which $\omega$ induces a
state in the obvious way. This state can be extended using the
Hahn--Banach theorem to a state $\widehat{\omega}$ on $\Af(\Nb)$ (Prop.~2.3.24
of~\cite{BratRob}), with the property $\widehat{\omega}(\alpha_\psi A)=\omega(A)$
for all $A\in\Af(\Mb)$. Of course there is no reason to suppose that
$\widehat{\omega}\in\Sf(\Nb)$, but, $\widehat{\omega}$ must be the
weak-$*$ limit of a sequence of states $\omega_n$ in $\Sf(\Nb)$ by Lemma~\ref{lem:Fell}.
This induces a sequence $\alpha_\psi^*\omega_n\in\Sf(\Mb)$ such that
\begin{equation*}
\alpha_\psi^*\omega_n(A) = \omega_n(\alpha_\psi A)\longrightarrow
\widehat{\omega}(\alpha_\psi A) = \omega(A)
\end{equation*}
for all $A\in\Af(\Mb)$. Thus $\omega$ belongs to the weak-$*$ closure of
$\alpha_\psi^*\Sf(\Nb)$. As $\omega$ was arbitrary the first part of the
result is proved. The second part follows because all nontrivial representations of
simple algebras are faithful. $\square$

We remark that this establishes local physical equivalence for
nontrivial locally covariant free field theories in which each
$\Af(\Mb)$ is a Weyl or CAR algebra. More generally, however, the lack
of an analogue to Fell's theorem for
general $*$-algebras renders local physical equivalence a nontrivial
addition to the structure of locally covariant quantum field theory. One
needs to check that---as one would expect---it does in fact hold for known
models: it is planned to address this elsewhere~\cite{FKV}. 

Our focus now reverts to quantum inequalities. For the rest of this
section, we will assume that $\Phi$ is a locally covariant field
associated with a locally covariant QFT $\Af$ and test space $\Df$. We
will then make a number of consistency checks on the definitions set out
above, each of which will turn out to use local physical
equivalence. In addition, we will show how this principle may be used
to infer constraints on AQIs on Minkowski space from information about
a family of spacetimes with toroidal spatial topology.

To begin, suppose one has a
sharp quantum inequality on each spacetime, for the largest class of
sampling functions possible. Is it necessarily locally
covariant? It would seem strange if covariance did not
favour the best possible bounds, and would suggest that our definitions
were defective. We analyse this issue for state-independent QIs.
Accordingly, suppose that $\Sf$ is a locally covariant state space for
the theory. Define
\begin{equation*}
\widetilde{\QQ}^a_\Mb(f) =-\inf_{\omega\in\Sf(\Mb)} \omega(\Phi_\Mb(f))
\end{equation*}
for each $f\in\Df(\Mb)$ and set $\QQ^a_\Mb(f)=\widetilde{\QQ}^a_\Mb(f)\II_{\Af(\Mb)}$ and
\begin{equation*}
\Ff(\Mb)=\{f\in\Df(\Mb):\widetilde{\QQ}^a_\Mb(f)<\infty\}\,,
\end{equation*}
which, by construction, provides a sharp state-independent absolute QI on the
largest class of sampling functions possible on each spacetime. (Of
course, for some fields $\Ff(\Mb)$ might be empty, or the bound might be trivial.)

We now proceed to analyse
the covariance properties of this quantum inequality, supposing that 
$\psi:\Mb\to\Nb$. Since each $\omega\in\Sf(\Nb)$ induces a 
state $\alpha_\psi^*\omega\in\Sf(\Mb)$, we have
\begin{eqnarray}
-\widetilde{\QQ}^a_\Mb(f) &\le& \inf_{\omega\in\Sf(\Nb)}
\alpha_\psi^*\omega(\Phi_\Mb(f)) = \inf_{\omega\in\Sf(\Nb)}
\omega(\alpha_\psi\Phi_\Mb(f))\nonumber\\
& =& \inf_{\omega\in\Sf(\Nb)}
\omega(\Phi_\Mb(\psi_*f))=
-\widetilde{\QQ}^a_\Nb(\psi_*f)\,,
\label{eq:QQineq}
\end{eqnarray}
from which we may conclude that $\psi_*\Ff(\Mb)\subset\Ff(\Nb)$ so we
have a morphism $\Ff(\psi)=\psi_*|_{\Ff(\Mb)}$ from $\Ff(\Mb)$ to $\Ff(\Nb)$ in $\Set$. It is
obvious that composition and the identity property hold, so in fact
$\Ff$ is a functor as required. 

However, $\QQ^a$ is not a natural transformation unless the inequality
in~\eqref{eq:QQineq} can be replaced by an equality. This hiatus may be
resolved provided that $\Sf$ respects local physical
equivalence. By definition, given any $\epsilon>0$ one may approximate
$-\widetilde{\QQ}^a_\Mb(f)$ to within $\epsilon/2$ by the expectation value of $\Phi_\Mb(f)$ in some state
$\omega\in\Sf(\Mb)$. Using local physical equivalence, $\omega(\Phi_\Mb(f))$ may
itself be approximated to within $\epsilon/2$ by $\alpha_\psi^*\omega'(\Phi_\Mb(f))$
for some state $\omega'\in\Sf(\Nb)$. But 
\begin{equation*}
\alpha_\psi^*\omega'(\Phi_\Mb(f))=\omega'(\Phi_\Nb(\psi_*f)) \ge 
-\widetilde{\QQ}^a_\Nb(\psi_*f)
\end{equation*}
using covariance and the QI on $\Nb$. Hence 
$-\widetilde{\QQ}^a_\Mb(f)+\epsilon\ge -\widetilde{\QQ}^a_\Nb(\psi_*f)$, and
since $\epsilon$ was arbitrary we may conclude, putting this together
with~\eqref{eq:QQineq}, that~\eqref{eq:Qt_trans}
now holds. Therefore $\QQ^a$ is natural and our absolute QI (though
possibly trivial) is locally covariant. We may formulate this as follows.

\begin{Prop}\label{prop:sharp_AQI} Suppose $\Phi$ is any locally covariant quantum field, associated with
the functors $\Af$ and $\Df$ and a state space $\Sf$ which respects
local physical equivalence. The sharp state-independent absolute QI on 
$\Phi$, relative to $\Sf$, defined on the largest class of
test functions possible on each globally hyperbolic spacetime, is automatically locally covariant. 
\end{Prop}

A sharp difference QI on $\Phi$ may be defined in a very
similar way, by setting  
\begin{equation*}
\widetilde{\QQ}_\Mb^{d}(f,\omega_0) 
= -\inf_{\omega\in\Sf(\Mb)}\left(\omega(\Phi_\Mb(f))-\omega_0(\Phi_\Mb(f))\right)\,.
\end{equation*}
It is obvious that this is defined on the same set $\Ff(\Mb)$ as the
absolute QI obtained above, and moreover that
\begin{equation*}
\widetilde{\QQ}_\Mb^{d}(f,\omega_0) = \widetilde{\QQ}_\Mb^{a}(f)+\omega_0(\Phi_\Mb(f))\,.
\end{equation*}
The reverse construction is also of interest. Suppose one is given a
locally covariant state-independent difference QI (not necessarily
sharp). Does there exist a locally covariant absolute QI on $\Phi$? The
answer to this is affirmative, subject to local physical equivalence:
first rearrange the basic difference inequality as
\begin{equation}
\widetilde{\QQ}_\Mb^d(f,\omega')-\omega'(\Phi_\Mb(f)) \ge -\omega(\Phi_\Mb(f))\,,
\label{eq:rearr}
\end{equation}
where $\omega'$ is the reference state and $\omega\in\Sf(\Mb)$ is
arbitrary. But now fix $\omega$ and allow $\omega'$ to vary. Since the 
left-hand side is clearly bounded from below, 
\begin{equation}
\widetilde{\QQ}_\Mb^a(f)\stackrel{\rm def}{=}\inf_{\omega'\in\Sf(\Mb)}\left(
\widetilde{\QQ}_\Mb^d(f,\omega')-\omega'(\Phi_\Mb(f))\right)
\label{eq:abs_from_diff}
\end{equation}
is finite for all $f\in\Ff(\Mb)$, and independent of $\omega$. Moreover, from~\eqref{eq:rearr}
\begin{equation*}
\omega(\Phi_\Mb(f)) \ge -\widetilde{\QQ}_\Mb^a(f)\,.
\end{equation*}
As $\omega$ was arbitrary, we obtain an absolute QI by setting
$\QQ^a_\Mb(f)=\widetilde{\QQ}^a_\Mb(f)\II$. 

To establish local covariance, we assume in addition that
$\omega\mapsto\widetilde{\QQ}_\Mb^d(f,\omega)$ is weak-$*$ continuous for
each $f$, and then proceed along lines similar to those used before. Because the
domain of sampling functions is unchanged, $\Ff$ has the required
properties; in addition we may calculate
\begin{eqnarray*}
\widetilde{\QQ}_\Mb^a(f) &\le& \inf_{\omega'\in\Sf(\Nb)}\left(
\widetilde{\QQ}_\Mb^d(f,\alpha_\psi^*\omega')-\alpha_\psi^*\omega'(\Phi_\Mb(f))\right)\\
&=&\inf_{\omega'\in\Sf(\Nb)}\left(
\widetilde{\QQ}_\Nb^d(\psi_*f,\omega')-\omega'(\Phi_\Nb(\psi_*f))\right)\\
&=&\widetilde{\QQ}_\Nb^a(\psi_*f)\,,
\end{eqnarray*}
so it is only necessary to show that the reverse inequality holds. This
proceeds by first approximating $\widetilde{\QQ}_\Mb^a(f)$ by some
$\widetilde{\QQ}_\Mb^d(f,\omega')-\omega'(\Phi_\Mb(f))$ and then using
local physical equivalence and weak-$*$ continuity of
$\omega'\mapsto\widetilde{\QQ}_\Mb^d(f,\omega')$ to approximate this, in
turn, by
$\widetilde{\QQ}_\Mb^d(f,\alpha_\psi^*\omega'')-\alpha_\psi^*\omega''(\Phi_\Mb(f))$
for some $\omega''\in\Sf(\Nb)$. The remainder of the argument 
runs parallel to that given above, and need not be
repeated. To summarise, we have established the following. 

\begin{Prop}\label{prop:DQI_to_AQI_by_inf} Suppose $\Phi$ is any locally covariant quantum field, associated with
the functors $\Af$ and $\Df$ and a state space $\Sf$ which respects
local physical equivalence. Let $\QQ^d$ be any state-independent locally
covariant difference QI on $\Phi$ relative to $\Sf$, with smearing
fields $\Ff$ such that
$\QQ^d_\Mb(f,\omega)=\widetilde{\QQ}_\Mb^d(f,\omega)\II_{\Af(\Mb)}$, with
$\omega\mapsto\widetilde{\QQ}_\Mb^d(f,\omega)$ weak-$*$ continuous for
each $f\in\Ff(\Mb)$ and every $\Mb$. Then
equation~\eqref{eq:abs_from_diff} defines a state-independent, locally covariant, absolute QI on $\Phi$,
relative to $\Sf$, with the same smearing fields $\Ff$. 
\end{Prop}

In Appendix~\ref{appx:Wick_square}, we will show that the hypothesis of weak-$*$ continuity
is satisfied for a DQI on the Wick square of the free scalar field, so
Prop.~\ref{prop:DQI_to_AQI_by_inf} applies. In this case, however, the
quantity inside the infimum in~\eqref{eq:abs_from_diff} is independent
of $\omega'$ (as is also shown in Appendix~\ref{appx:Wick_square}), so the bound obtained
coincides with that obtained by rearrangement in Sect.~\ref{sect:DQIs}.
The Wick square DQI also obeys the hypotheses of part (b) of the
following result, which demonstrates
that our notion of a (non)trivial QI is compatible with local covariance.
\begin{Prop}\label{prop:covariance_of_triviality}
Consider a locally covariant quantum field $\Phi$
associated with a theory respecting local physical equivalence.
(a) Let $\QQ^a:\Ff\nto\Af$ be a locally covariant absolute QI on $\Phi$
relative to $\Sf$. Suppose $\psi:\Mb\to\Nb$ in $\Man$. If $\QQ^a$ is
trivial on $\Nb$ then it is trivial on $\Mb$. (Conversely, if $\QQ^a$ is
nontrivial on $\Mb$ then it is nontrivial on $\Nb$.) (b) Suppose 
$\QQ^d:\langle\Ff,\Sf^\op\rangle\nto\Af$ is a locally covariant DQI
such that $\Sf(\Mb)\times\Sf(\Mb)\owns (\omega,\omega_0)\mapsto \omega(\QQ_\Mb(f,\omega_0))$ is weak-$*$
continuous in $\omega_0$, uniformly in $\omega$, for each
$f\in\Ff(\Mb)$ and every $\Mb\in\Man$. If $\psi:\Mb\to\Nb$ in $\Man$ and $\QQ^d$
is trivial on $\Nb$, then $\QQ^d$ is trivial on $\Mb$.
\end{Prop}
{\em Proof:} (a) Choose any fixed
positive constant $c$ with the dimensions of
$\omega(\QQ^a_\Mb(f))$. Then triviality on any spacetime $\Mb$
is equivalent to the statement that
\begin{equation}
\sup_{\omega\in\Sf(\Mb)}
\frac{|\omega(\Phi_\Mb(f))|}{|\omega(\QQ^a_\Mb(f))|+c}
<\infty \label{eq:AQI_triv}
\end{equation}
for all $f\in\Ff(\Mb)$. Supposing that $\QQ$ is trivial on $\Nb$, we may
use covariance to observe that
$\omega\mapsto |\omega(\Phi_\Mb(f))|/(|\omega(\QQ^a_\Mb(f))|+c)$ is bounded on the set
$\alpha_\psi^*\Sf(\Nb)$. Now take an arbitrary $\omega\in\Sf(\Mb)$;
using local physical equivalence we may find a sequence
$\omega_n\in\alpha_\psi^*\Sf(\Nb)$ such that
$\omega_n(\Phi_\Mb(f))\to\omega(\Phi_\Mb(f))$ and
$\omega_n(\QQ_\Mb(f))\to\omega(\QQ_\Mb(f))$. Combining this with our
first observation we see that \eqref{eq:AQI_triv} holds, so $\QQ^a$ is
trivial on $\Mb$. 

The proof of part (b) is similar. Choosing any fixed
constant $c$ with the dimensions of
$\omega(\QQ_\Mb(f,\omega_0))$, triviality on any spacetime $\Mb$
is equivalent to the statement that
\begin{equation*}
\sup_{\omega\in\Sf(\Mb)}
\frac{|\omega(\Phi_\Mb(f))|}{|\omega(\QQ^d_\Mb(f,\omega_0))|+c}
<\infty
\end{equation*}
for all $\omega_0\in\Sf(\Mb)$ and $f\in\Ff(\Mb)$. Using triviality on
$\Nb$, we may infer that the ratio
$|\omega(\Phi_\Mb(f))|)/(|\omega(\QQ^d_\Mb(f,\omega_0))|+c)$ is bounded
for all $\omega\in\alpha_\psi^*\Sf(\Nb)$ for each fixed $\omega_0\in\alpha_\psi^*\Sf(\Nb)$. We use local physical
equivalence to extend this to all $\omega\in\Sf(\Mb)$, and then the
uniform weak-$*$ continuity hypothesis and local physical equivalence
to extend again to all $\omega_0\in\Sf(\Mb)$. Hence $\QQ^d$ is trivial
on $\Mb$. $\square$

To conclude this section,
we show how information on a class of spacetimes with toroidal spatial
topology can be used to infer information about AQIs in Minkowksi
space.  Consider a theory $\Af:\Man\to\TAlg$ with a
state space $\Sf$ which respects local physical equivalence, and let $\Phi:\Df\nto\Af$ be a
field of the theory. To keep matters simple, we restrict to the
situation where $\Df(\Mb)=C_0^\infty(\Mb)$. We assume in addition 
that each $\Phi_\Mb$ is a linear map, that
$\Phi_\Mb(f)^*=\Phi_\Mb(\overline{f})$ for all $f\in\Df(\Mb)$ and
that the one-point functions of $\Phi$ with respect to $\Sf$ are
smooth, i.e., for each $\omega\in\Sf(\Mb)$ there is a smooth function
$p\mapsto \omega(\Phi_\Mb(p))$ on $\Mb$ such that 
\begin{equation*}
\omega(\Phi_\Mb(f))=\int_\Mb \omega(\Phi_\Mb(p)) f(p) \,d{\rm vol}_\Mb(p)
\end{equation*}
for all $f\in\Df(\Mb)$ and each $\Mb\in\Man$. 

Let $\Mb_0$ be $n$-dimensional Minkowski space, and fix a system of inertial
coordinates $(t,x_1,\ldots,x_{n-1})$. Define $\Nb_L$ to be the
quotient of $\Mb_0$ by the group $\ZZ^{n-1}$ generated by 
translations through proper distances $L$ along
the $x_i$-axes, so each $\Nb_L$ has the topology of
$\RR\times\TT^{n-1}$. Suppose that each $\Sf(\Nb_L)$ contains a
set of translationally invariant states $S_L$ and set
\begin{equation*}
\kappa(L)=  \inf_{\omega\in S_L} \omega(\Phi_{\Nb_L}(p))
\end{equation*}
(which is independent of the particular $p\in\Nb_L$, of course). 
In the typical situation of interest for QIs, the function $\kappa$ is
monotone increasing in $L$ and $\kappa(L)\to -\infty$ as $L\to
0^+$. We will assume this in what follows. One final definition: given
any subset $S$ of $\Mb_0$, the {\em timelike diameter} of $S$ is the
infimum, over all points $p,q$ for which
$S\subset I^+(p)\cap I^-(q)$, of the interval between $p$ and $q$.  

\begin{Prop} For each nonnegative $f\in\Df(\Mb_0)$, we have
\begin{equation*}
\inf_{\omega\in\Sf(\Mb_0)} \omega(\Phi_{\Mb_0}(f)) \le
\kappa(\ell(f))\int_{\Mb_0} f \,d{\rm vol}_{\Mb_0}\,,
\end{equation*}
where $\ell(f)$ is the timelike diameter of the support of $f$. 
\end{Prop}
{\noindent\em Proof:} Let $p_\pm\in\Mb_0$ be any points such that
\begin{equation*}
\supp f \subset D\stackrel{\rm def}{=} I_{\Mb_0}^+(p_-)\cap I_{\Mb_0}^-(p_+)
\end{equation*}
and construct inertial coordinates $x'=(t',\xb')$ in which $p_\pm$
have coordinates $(\pm L/2,\Ob)$, so that $D$ is described by the
inequality $|t'|+|\xb|<L/2$. Endowing $D$ with the metric and
(time)-orientation induced from $\Mb_0$, it becomes an object $\Db$ of
$\Man$ in its own right, with a morphism $\iota:\Db\to\Mb_0$ given by
the inclusion mapping. In addition there is a morphism
$\psi:\Db\to\Nb_L$, which is the ``smallest'' of the $\Nb_{L'}$ into
which $\Db$ may be embedded (note that $\Db$ is open, so it is just
too small to detect the topology of $\Nb_L$). 

Given any $\omega\in S_L$, the state $\alpha_\psi^*\omega\in\Sf(\Db)$ obeys
\begin{eqnarray*}
\alpha_\psi^*\omega(\Phi_\Db(\iota^* f)) &=&
\omega(\Phi_{\Nb_L}(\psi_*\iota^*f))
=\omega(\Phi_{\Nb_L}(p))\int_{\Nb_L} \psi_*\iota^*f\,d{\rm
vol}_{\Nb_L} \\
&=&\omega(\Phi_{\Nb_L}(p))\int_{\Mb_0} f\,d{\rm
vol}_{\Mb_0}\,,
\end{eqnarray*}
where $p\in\Nb_L$ is arbitrary. Given an arbitrary $\epsilon>0$,
therefore, there exists a state $\omega'\in\Sf(\Db)$ such that
\begin{equation*}
\omega'(\Phi_\Db(\iota^* f))< (\kappa(L)+\epsilon)\int_{\Mb_0} f\,d{\rm
vol}_{\Mb_0}
\end{equation*}
and, by local physical equivalence, there exists $\omega''\in\Sf(\Mb_0)$ such that
\begin{equation*}
|\omega''(\Phi_{\Mb_0}(f))-\omega'(\Phi_\Db(\iota^*f))|<\epsilon\int_{\Mb_0} f\,d{\rm
vol}_{\Mb_0}\,.
\end{equation*}
Since all the expectation values are real and $f$ is nonnegative, we have
\begin{equation*}
\omega''(\Phi_{\Mb_0}(\iota^* f))< (\kappa(L)+2\epsilon)\int_{\Mb_0} f\,d{\rm
vol}_{\Mb_0}
\end{equation*}
and so, taking $\epsilon\to 0^+$ and also optimising over all double
cones containing $\supp f$, we obtain the required result. $\square$

It follows immediately that any state independent QI on $\Phi$,
relative to the nonnegative functions in $\Df(\Mb)$, must obey
\begin{equation*}
\widetilde{\QQ}^a_{\Mb_0}(f) \ge -\kappa(\ell(f))\int_{\Mb_0} f\,d{\rm
vol}_{\Mb_0}\,,
\end{equation*}
so the scaling behaviour of $\kappa$ bounds that of $\widetilde{\QQ}^a_{\Mb_0}(f)$
if one shrinks the support of $f$ while holding its integral
constant. A similar situation occurs if the support of $f$ 
converges to a null geodesic segment with endpoints $q_\pm$ (with
$q_-$ to the past of $q_+$) because the double cones spanned by
$p_\pm=\tau_{\pm\epsilon}q_\pm$ have timelike diameter of order
$\epsilon$ as $\epsilon\to 0^+$, where $\tau_s$ is translation along a constant future-pointing
timelike vector field. Hence $\widetilde{\QQ}^a_{\Mb_0}(f_n)\to\infty$
for any sequence $f_n$ with $\int_{\Mb_0}f_n\,d{\rm vol}_{\Mb_0}$
fixed for all $n$, whose support shrinks onto a null geodesic segment.
This is consistent with the facts that there are no QEIs for
the free scalar field for averaging along finite portions of null
curves~\cite{Fe&Ro03} and (not unrelated) that there are not expected to be
nontrivial observables localised on null geodesic segments in
spacetimes of dimensions $d\ge 2$ (see \cite{Woronowicz} and footnotes
34 and 35 in~\cite{Fe&Ro03}).

\section{The covariant numerical range and spectrum}\label{sec:order}

Let $\Df$ be any covariant set of smearing fields and $\Af$ a locally
covariant quantum field theory. The set\footnote{See below for a proof
that this is in fact a small set.} $\Nat(\Df,\Af)$ of natural
transformations, i.e., those quantum fields of the model whose smearings
are drawn from $\Df$, may be given a partial order as follows: 
we write $\Phi\ge \Psi$ if
\begin{equation}
\omega(\Phi_\Mb(f)) \ge \omega(\Psi_\Mb(f))\quad \forall
\Mb\in\Man,~f\in\Df(\Mb),~\omega\in\Sf(\Mb)\,.
\label{eq:order}
\end{equation}
Of course, if $\Df(\Mb)$ is a vector space, and $\Phi_\Mb$ and $\Psi_\Mb$ are linear maps from $\Df(\Mb)$
to $\Af(\Mb)$, it cannot hold that $\Phi\ge\Psi$ unless $\Phi$ and
$\Psi$ are {\em $\Sf$-indistinguishable}, i.e.,
$\omega(\Phi_\Mb(f))=\omega(\Psi_\Mb(f))$ for all $\Mb\in\Man$,
$f\in\Df(\Mb)$ and $\omega\in\Sf(\Mb)$, because the inequality in
\eqref{eq:order} must hold for both $f$ and $-f$. Recall, however, that we do not
require in general that fields are linear or that sets of smearing
functions are vector spaces (this applies in particular to the
restricted sets of smearing functions considered in QIs). 

Quantum Inequalities can be succinctly formulated using order relations
of this type. Thus, an absolute QI on $\Phi\in\Nat(\Df,\Af)$ may be expressed in terms of a
subfunctor $\Ff\subset\Df$ and a field $\QQ^a\in\Nat(\Ff,\Af)$ such
that
\begin{equation*}
\Phi|_\Ff \ge - \QQ^a
\end{equation*}
with respect to the order relation on $\Nat(\Ff,\Af)$,
where $-\QQ^a$ has components $(-\QQ^a)_\Mb(f) =
-\QQ^a_\Mb(f)$. Note that in all QEIs currently known, $\Ff(\Mb)$ is a convex
subset of a vector space, but not a vector space in its own right, and
$f\mapsto\QQ^a_\Mb(f)$ is nonlinear. Similarly, a DQI on $\Phi$ may be written in the form
\begin{equation*}
\Delta\Phi|_\Ff \ge - \QQ^d\,,
\end{equation*}
where $\QQ^d\in\Nat(\langle\Ff,\Sf^\op\rangle,\Af)$ and
$\Delta\Phi\in\Nat(\langle\Df,\Sf^\op\rangle,\Af)$ is defined by
\begin{equation*}
(\Delta \Phi)_\Mb(f,\omega)= \Phi_\Mb(f) -
\omega(\Phi_\Mb(f))\II_{\Af(\Mb)}
\end{equation*}
and is natural owing to the identity
\begin{eqnarray*}
(\Delta
\Phi)_\Nb(\psi_*f,\omega)&=&\Phi_\Nb(\psi_*f)-\omega(\Phi_\Nb(\psi_*f))\II_{\Af(\Nb)}
\nonumber\\
&=&
\alpha_\psi(\Phi_\Mb(f)-\alpha_\psi^*\omega(\Phi_\Mb(f))\II_{\Af(\Mb)}) \\
&=&\alpha_\psi((\Delta\Phi)_\Mb(f,\alpha_\psi^*\omega))
\end{eqnarray*}
holding whenever $\psi:\Mb\to\Nb$ in $\Man$, for all $f\in\Df(\Mb)$
and $\omega\in\Sf(\Nb)$.

In the rest of this section, we will set order relations of this type
in a broader context, which will lead naturally to notions of a
numerical range and spectrum of a locally covariant
field, and also to an algebra of fields abstracted from particular
spacetimes or smearings. This appears to provide a new way to analyse
locally covariant fields, in which all constructions are automatically
natural (i.e., covariant). The discussion of numerical range remains
reasonably close to the subject of quantum inequalities; the spectrum is
perhaps less immediately relevant, but its elementary theory is
developed to demonstrate that it has the usual relationship to numerical
range, and to illustrate the potential for covariant functional calculus
of quantum fields. 

Before we embark on this, it is necessary to dispose of a
set-theoretical problem. The category $\Man$ is a large category: that
is, $\obj\Man$ is not a small set\footnote{Given a manifold $M$,
any set isomorphic to $M$ (as a set) may be given the structure of a
manifold by utilising the isomorphism, whereupon it is isomorphic to $M$
as a manifold. Accordingly, $\obj\Man$ must be at least as large as the 
set of all small sets of cardinality $\aleph_c$, and is therefore a large set.}.
The Whitney embedding theorem provides the standard solution to this
problem, as it asserts that every smooth manifold of dimension $d$ may be embedded as a
smooth submanifold of $\RR^{2d+1}$. Thus the collection of isomorphism
equivalence classes in the category of smooth
manifolds of dimension $d$ may be identified with a subset of the power
set of $\RR^{2d+1}$, and is therefore a small set. The objects of $\Man$
are distinguished further by their metrics and (time-)orientations but,
as it is easy to see that these are drawn from small sets, the argument extends straightforwardly
to show that the isomorphism equivalence classes in
$\Man$ form a small set, i.e., $\Man$ is an essentially small category. 
We now choose one representative from each
isomorphism class, to obtain a small set $\Mfr$ of `basic spacetimes'.
The precise choice of these representatives will never be important. 

Now note that any natural transformation $\Phi:\Df\nto\Af$ is completely
determined by its components $\Phi_\Mb$ for $\Mb\in\Mfr$. For each
$\Mb\in\Man$ there is a unique $\widetilde{\Mb}\in\Mfr$ for which 
$\Mb\stackrel{\psi}{\longrightarrow}\widetilde{\Mb}$, with $\psi$ a $\Man$-isomorphism.
Thus $\alpha_\psi:\Af(\Mb)\to\Af(\widetilde{\Mb})$ is a
$\TAlg$-isomorphism and 
$\Phi_\Mb=\alpha_\psi^{-1}\circ\Phi_{\widetilde{\Mb}}\circ \psi_*$
because $\Phi$ is natural. It
follows that $\Nat(\Df,\Af)$ is a small set, as it is isomorphic to a subset of the
Cartesian product over $\Mb\in\Mfr$ of the set of functions from $\Df(\Mb)$ to
$\Af(\Mb)$.

\subsection{Numerical range}

We begin by defining the numerical range relative
to a given state space. If $\Aa$ is a topological $*$-algebra and
$S\subset \Aa^*_{+,1}$ is convex, we define the numerical
range of $A\in\Aa$, relative to $S$, by 
\begin{equation*}
N_{\Aa,S}(A) = \cl\{\omega(A): \omega\in S\}\,,
\end{equation*}
where $\cl$ denotes the closure in the topology of $\CC$. 
Owing to convexity of $S$, $N_{\Aa,S}(A)$ is convex for all
$A$.

The numerical range is a well-known tool in the theory of quadratic forms
in Hilbert spaces (and, in particular, matrix theory) \cite{GustafsonRao}. A
corresponding theory in Banach-$*$-algebras is described in
\cite{BonsallDuncan} (see also~\cite{Palmer_vol1}). 
We have generalised this to $\TAlg$ and permitted
a restricted space of states; in addition, we differ from the
references mentioned by taking the closure in our definition.
This is convenient in our case, as shown by the following lemma.
 
\begin{Lem} \label{lem:nr_natural}
Suppose $\Aa,\Ba\in\TAlg$. If $\alpha:\Aa\to\Ba$ is a faithful, unit-preserving
$*$-homomorphism and $T\subset \Ba^*_{+,1}$ obeys (i)
$\alpha^*T\subset S$ and (ii) $\alpha^* T$ has the same weak-$*$
closure as $S$ in $\Aa^*_{+,1}$, then we have 
\begin{equation*}
N_{\Ba,T}(\alpha(A)) = N_{\Aa,S}(A)
\end{equation*}
for all $A\in\Aa$. 
\end{Lem}
{\noindent\em Proof:} Fix $A\in\Aa$. Given any $\omega\in T$, it is clear that 
$\omega(\alpha(A))=\alpha^*\omega(A)\in N_{\Aa,S}(A)$ and hence
that the left-hand side is contained in the right. On the other hand,
choose any $\omega\in S$; then there exists a sequence $\omega_n\in T$
such that $\omega_n(\alpha(A))=\alpha^*\omega_n(A)\to \omega(A)$,
from which the reverse inclusion follows. $\square$

Now return to locally covariant QFT $\Af$, with states $\Sf$
satisfying local physical equivalence. Writing
$2^\CC:\Man\to\Set$ for the constant functor that assigns
the power set of $\CC$ to each $\Mb\in\Man$ and the identity morphism $\id_{2^\CC}$
to each morphism of $\Man$, Lemma~\ref{lem:nr_natural} entails that
the maps $N_\Mb(A)=N_{\Af(\Mb),\Sf(\Mb)}(A)$ constitute a natural
transformation $N:\Af\nto 2^\CC$ expressed by commutativity of the diagram
\begin{equation*}
\begin{CD}
\Mb @.\phantom{AAAA}@.\Af(\Mb) @>N_{\Mb}>> 2^\CC \\
@V{\psi}VV @. @V{\alpha(\psi)}VV     @VV{\id_{2^\CC}}V\\
\Nb @.\phantom{AAAA}@.\Af(\Nb)@>>N_{\Nb}> 2^\CC
\end{CD}\qquad.
\end{equation*}
Moreover, if we have a field $\Phi:\Df\nto\Af$, then $N(\Phi)_\Mb(f)=N_\Mb(\Phi_\Mb(f))$
defines a natural transformation $N(\Phi):\Df\nto 2^\CC$. 

We may use the numerical range to rephrase the construction of the sharp
state-independent AQI of Prop.~\ref{prop:sharp_AQI}: we have
$\widetilde{\QQ}^a_\Mb(f)=-\inf N(\Phi)_\Mb(f)$, which we could write as
$\widetilde{\QQ}^a=\inf N(\Phi)$. 

For future reference, we note the following. If
$\psi:\Mb\to\Nb$ is an isomorphism in $\Man$, then 
\begin{equation}
\bigcup_{f\in\Df(\Mb)}
N(\Phi)_\Mb(f)
= \bigcup_{f\in\Df(\Mb)}
N(\Phi)_\Nb(\psi_*f)
= \bigcup_{f\in\Df(\Nb)} N(\Phi)_{\Nb}(f)\,,
\label{eq:invariant_numerical_range}
\end{equation}
where we have used the fact that $N(\Phi)$ is natural and the fact that
$\psi_*:\Df(\Mb)\to\Df(\Nb)$ is an isomorphism. Accordingly, these sets
are spacetime invariants. 

The order relation \eqref{eq:order} may now be expressed as follows:
$\Phi\ge 0$ if and only if $N(\Phi)_\Mb(f)\subset[0,\infty)$ for all
$\Mb\in\Man$ and $f\in\Df(\Mb)$, and $\Phi\ge \Psi$ when
$\Phi-\Psi\ge 0$. Here $\Phi-\Psi$ has components $(\Phi-\Psi)_\Mb(f)=
\Phi_\Mb(f)-\Psi_\Mb(f)$ and is obviously a natural transformation
between $\Df$ and $\Af$. 

\begin{Lem} $\Phi\ge 0$ if and only if
\begin{equation*}
\nu(\Phi)\stackrel{\rm def}{=} \cl\,\co\left(\bigcup_{\Mb\in\Mfr}\bigcup_{f\in\Df(\Mb)}
N(\Phi)_\Mb(f)\right)
\end{equation*}
is contained in $[0,\infty)$, where $\co$ is the operation
of forming the convex hull. 
\end{Lem}
{\noindent\bf Remark:} In Mac~Lane's description of category theory
founded on a single universe~\cite{MacLane69}, it is permissible to
index a union over a small set, which is why we have used $\Mfr$ instead
of the large set $\obj\Man$. It follows
from the proof that $\nu(\Phi)$ is independent of the particular choice of
basic manifolds. In fact, the set-theoretical problem is not at all severe, because all
the sets in the union are subsets of $\CC$, so we could write
\begin{equation*}
\nu(\Phi) = \cl\,\co\left\{z\in \CC: \exists \Mb\in\Man,~f\in\Df(\Mb)~{\rm s.t.}~z=N(\Phi)_\Mb(f)\right\}\,,
\end{equation*}
which is a legitimate subset selection within ZFC (see, e.g., Sect.~I.5 of~\cite{Kunen}). 
However, it is convenient to be able to use the union notation freely
without abuse. 

{\noindent\em Proof:} As $[0,\infty)$ is closed and convex it is enough
to check that $\Phi\ge 0$ if and only if the union in parentheses is
contained in $[0,\infty)$. But
\eqref{eq:invariant_numerical_range} and the fact that $\Mfr$ contains
a representative of every isomorphism class in $\Man$, show that this is
equivalent to the condition that $N(\Phi)_\Mb(f)\subset [0,\infty)$
for all $\Mb\in\Man$, $f\in\Df(\Mb)$, which is the condition that
$\Phi\ge 0$. $\square$

Note that $\nu(\Phi)$ cannot expand, and
may contract, if $\Df$ is replaced by one of its subfunctors. Indeed, in
many circumstances it may be necessary to make a replacement like this
in order to cut the numerical range down from all of $\CC$ or $\RR$. 

In the previous result, the formation of the closed convex hull was redundant,
but defining $\nu(\Phi)$ in the above
way has the advantage that $\nu(\Phi)$ may be regarded as a numerical range in its
own right. Notice that the set of natural transformations
$\Nat(\Df,\Af)$ may be given the structure of a $*$-algebra, with sums
and products defined pointwise, i.e.,
\begin{eqnarray*}
\Phi^*_\Mb(f)&=& (\Phi_\Mb(f))^* \\
(\lambda\Phi+\mu\Psi)_\Mb(f) &=& \lambda\Phi_\Mb(f) + \mu\Psi_\Mb(f)\\
(\Phi\Psi)_\Mb(f) &=& \Phi_\Mb(f)\Psi_\Mb(f)\,,
\end{eqnarray*}
which are clearly both associative and distributive. There is a unit $\II:\Df\nto\Af$ with components
$\II_\Mb(f)=\II_{\Af(\Mb)}$, and we may endow the algebra with the
topology of pointwise convergence, i.e., a net $\Phi_\alpha$ converges
to $\Psi$ if we
have $\Phi_{\alpha\,\Mb}(f)\to\Psi_\Mb(f)$ for all $\Mb\in\Man$,
$f\in\Df(\Mb)$. We denote the resulting unital topological $*$-algebra
by $\Fa(\Df,\Af)$. 

Next, observe that each $(\Mb,f,\omega)\in
\Man\times\Df(\Mb)\times\Sf(\Mb)$ induces a linear functional
$\xi_{\Mb,f,\omega}$ on $\Fa(\Df,\Af)$ by
\begin{equation*}
\xi_{\Mb,f,\omega}(\Phi) = \omega(\Phi_\Mb(f))\,,
\end{equation*}
which is clearly continuous in $\Phi$, positive
[$\xi_{\Mb,f,\omega}(\Phi^*\Phi)=\omega(\Phi_\Mb(f)^*\Phi_\Mb(f))\ge
0$] and normalised 
[$\xi_{\Mb,f,\omega}(\II)=\omega(\II_{\Af(\Mb)})=1$]. If we define
$S(\Df,\Af)$ to consist of all finite convex combinations of states of
this type, it is then immediate that
\begin{equation*}
\nu(\Phi) = N_{\Fa(\Df,\Af),S(\Df,\Af)}(\Phi)\,.
\end{equation*}
Thus, a field $\Phi$ is positive, i.e., $\Phi\ge 0$, if and only if its numerical range in
$\Fa(\Df,\Af)$, relative to state space $S(\Df,\Af)$ is contained in
$[0,\infty)$. Note that the state space $S(\Df,\Af)$
contains states which are mixtures of states
associated with the theory on {\em different} spacetimes.

An important question is under what circumstances the infimum of the
numerical range of a field $\Phi$ is attained. The following result shows that this
cannot be the case [except for trivial situations] for any state $\omega$ which is separating for linear
combinations of $\Phi$ and the identity field, in the sense that 
$\omega((\Phi_\Mb(f)+\mu\II_{\Af(\Mb)})^*(\Phi_\Mb(f)+\mu\II_{\Af(\Mb)}))=0$
for some $\mu\in\CC$ implies that
$\Phi_\Mb(f)=-\mu\II_{\Af(\Mb)}$. In situations where a separating
vacuum state exists (e.g., from a Reeh--Schlieder property) then the
result shows that there must be states with expectation values for
$\Phi_\Mb$ below that of the vacuum state. The argument is based on the proof
of Lemma 1 of \cite{EGJ}. 

\begin{Prop} Suppose $\nu(\Phi)\subset[\nu_0,\infty)$, and that $\Sf$ is
closed under operations induced by $\Af$. Suppose further that
$\omega\in\Sf(\Mb)$ is separating for linear combinations of $\Phi$
and the identity field. If
$\Phi_\Mb(f)=\Phi_\Mb(f)^*$ obeys $\omega(\Phi_\Mb(f))=\nu_0$, then $\Phi_\Mb(f)=\nu_0\II_{\Af(\Mb)}$. 
\end{Prop}
{\noindent\em Proof:} The field $\Psi=\Phi-\nu_0\II$ is positive, so
we have a semidefinite sesquilinear
form $(A,B)\mapsto \omega(A^*\Psi_\Mb(f)B)$ and hence a
Cauchy--Schwarz inequality
\begin{equation*}
|\omega(A^*\Psi_\Mb(f)B)|^2 \le
\omega(B^*\Psi_\Mb(f)B)\omega(A^*\Psi_\Mb(f)A) \,.\label{eq:CSmod}
\end{equation*}
Setting $B=\II_{\Af(\Mb)}$ and $A=\Psi_\Mb(f)$, we deduce that
$\omega(\Psi_\Mb(f)^*\Psi_\Mb(f))=0$ (because $\omega(\Psi_\Mb(f)=0$) and hence that
$\Phi_\Mb(f)=\nu_0\II_{\Af(\Mb)}$ using the separating property.
$\square$ 

The algebra $\Fa(\Df,\Af)$ is of interest in its own right. It
consists of the locally covariant fields of the theory, but abstracted
from particular smearings in particular spacetimes (by virtue of
knowing about all possible smearings in all possible
spacetimes). Constructions conducted in this algebra and related
structures are
automatically natural -- a point which we will develop in more detail
for theories described by $C^*$-algebras. 
The following result is a
consistency check on the `naturalness' of the construction of
$\Fa(\Df,\Af)$. 

\begin{Prop} Suppose $(\Df_i,\Af_i)$, for $i=1,2$, are equivalent in
the sense that there are natural transformations
$\alpha:\Af_1\nto\Af_2$ and $\delta:\Df_1\nto\Df_2$ with each
$\alpha_\Mb$ and $\delta_\Mb$ an isomorphism in $\TAlg$ and $\Set$,
respectively. Then each $\Phi\in\Nat(\Df_1,\Af_1)$ induces a natural
transformation $\alpha\circ\Phi\circ\delta^{-1}:\Df_2\nto\Af_2$, and
the map $\iota_{\delta,\alpha}\Phi\mapsto \alpha\circ\Phi\circ\delta^{-1}$ is an
isomorphism of $\Fa(\Df_1,\Af_1)$ and $\Fa(\Df_2,\Af_2)$ in $\TAlg$.
\end{Prop}
{\noindent\em Proof:} Compositions of natural transformations are natural, so
$\iota_{\delta,\alpha}(\Phi)\in \Fa(\Df_2,\Af_2)$ for each
$\Phi\in\Fa(\Df_1,\Af_1)$. 
The fact that $\iota_{\delta,\alpha}$ respects the $*$-algebraic
operations and preserves the unit follows from the fact that each
$\alpha_\Mb$ is a $*$-homomorphism. Continuity holds because
$\Phi_\nu\to\Psi$ implies that $(\iota_{\delta,\alpha}\Phi_\nu)_\Mb(f)
=\alpha_\Mb(\Phi_{\nu\,\Mb}(\delta_\Mb^{-1}(f)))\to
\alpha_\Mb(\Psi_\Mb(\delta_{\Mb}^{-1}(f)))=(\iota_{\delta,\alpha}\Psi)_\Mb(f)$
for all $\Mb\in\Man$ and $f\in\Df_2(\Mb)$, so
$\iota_{\delta,\alpha}\Phi_\nu\to\iota_{\delta,\alpha}\Psi$. 
Since $\iota_{\delta,\alpha}$ has
the obvious inverse $\iota_{\delta^{-1},\alpha^{-1}}$ with the same
properties, we conclude that it is a $\TAlg$ isomorphism. $\square$

As a digression, we mention that there are other possible algebraic
combinations of fields. In particular, one may define a bi-local product
$\odot$, mapping $(\Phi,\Psi)\in\Nat(\Df,\Af)\times \Nat(\Df,\Af)$ to
$\Phi\odot\Psi\in \Nat(\langle\Df,\Df\rangle,\Af)$ such that
\begin{equation*}
(\Phi\odot\Psi)_\Mb(f,g) = \Phi_\Mb(f)\Psi_\Mb(g)
\end{equation*}
and similarly bi-local sums and $n$-local sums and products. The
resulting algebraic structures seem to offer compact, manifestly
natural, expressions of commutation relations, and might be worthy 
of further study.  

\subsection{Spectrum}

We now specialise to the case of theories where each $\Af(\Mb)$ is a
$C^*$-algebra and each $\alpha_\psi$ is a faithful, unit-preserving
$C^*$-morphism. In this context it is natural to restrict to a
$*$-subalgebra of $\Fa(\Df,\Af)$, which we denote
$\Fa^\infty(\Df,\Af)$, consisting of those
$\Phi\in\Fa(\Df,\Af)$ for which 
\begin{equation*}
\|\Phi\| \stackrel{\rm def}{=} \sup_{\Mb\in\Mfr}\sup_{f\in\Df(\Mb)}
\|\Phi_\Mb(f)\|_{\Af(\Mb)}
\end{equation*}
is finite. Note that this is independent of the choice of basic
spacetimes in $\Mfr$, because the inner supremum is constant on any isomorphism class in
$\Man$. 

\begin{Prop} 
$\Fa^\infty(\Df,\Af)$ is a $C^*$-algebra when equipped
with the norm $\|\cdot\|$. 
\end{Prop}
{\noindent\em Proof:} If is enough to check that $\Fa^\infty(\Df,\Af)$
is complete and that $\|\cdot\|$ has the $C^*$-property, as it is
clear that $\Fa^\infty(\Df,\Af)$ is a $*$-algebra. To check
completeness, note that any Cauchy sequence
$\Phi_n$ in $\Fa^\infty(\Df,\Af)$ induces Cauchy sequences
$\Phi_{n\,\Mb}(f)$ in $\Af(\Mb)$ for each $\Mb\in\Man$,
$f\in\Df(\Mb)$. Denoting the corresponding limit by $\Phi_{\Mb}(f)$,
one need only check that $\alpha_\psi(\Phi_\Mb(f)) = \Phi_{\Nb}(\psi_*
f)$ (using continuity of $\alpha_\psi$) to show that
$\Phi_{\Mb}:f\mapsto \Phi_\Mb(f)$ form the components of a natural
transformation $\Phi:\Df\nto\Af$. As $\Phi_n$ is Cauchy, there
exists $m$ such that
$\|\Phi_{n\,\Mb}(f)-\Phi_{m\,\Mb}(f)\|_{\Af(\Mb)}<1$ for all
$\Mb\in\Man$, $f\in\Df(\Mb)$ and $n\ge m$. From this it follows (taking
$n\to\infty$) that
$\|\Phi_{\Mb}(f)\|\le 1+\|\Phi_{m\,\Mb}(f)\|\le 1+\|\Phi_m\|$, so
$\Phi\in\Fa^\infty(\Df,\Af)$. Again, the Cauchy property for $\Phi_n$
implies that the convergence $\Phi_{n\,\Mb}(f)\to\Phi_\Mb(f)$ occurs uniformly 
in $\Mb$ and $f$, so $\Phi_n\to\Phi$ in $\Fa^\infty(\Df,\Af)$. The
$C^*$-property follows straightforwardly from the $C^*$-property of
each $\|\cdot\|_{\Af(\Mb)}$.  
$\square$

In many circumstances, it may be necessary to replace $\Df$ by one of
its subfunctors in order to obtain a nontrivial algebra. Thus, for
example, we might restrict to the unit ball with respect to a semi-norm
on $\Df(\Mb)$, to keep the supremum over $\Df(\Mb)$ bounded for certain
fields of interest. As an example of a nontrivial algebra
$\Fa^\infty(\Df,\Af)$, let $\Df(\Mb)=\CoinX{\Mb}$ and $\Af$ be the theory
consisting of Weyl algebras of the free scalar field. Then
$\Fa^\infty(\Df,\Af)$ clearly contains the `Weyl field'
$W:\Df\nto\Af$, defined so that $W_\Mb(f)$ is the Weyl generator associated with
the test function $f\in\Df(\Mb)$ [informally,
$W_\Mb(f)=e^{i\varphi_\Mb(f)}$].

The numerical range of fields in $\Fa^\infty(\Df,\Af)$ may be defined
as before, relative to the state space $S(\Df,\Af)$ [which is also a
state space for $\Fa^\infty(\Df,\Af)$]. But we can now also invoke the
spectrum, which is guaranteed to be well-behaved in the $C^*$-setting.

Let $\Sp_\Aa(A)$ denote the spectrum of an element $A$ of $C^*$-algebra
$\Aa$. If $\alpha:\Aa\to\Ba$ is a
unit-preserving faithful $*$-homomorphism between $C^*$-algebras $\Aa$
and $\Ba$, then $\Sp_\Ba(\alpha(A))=\Sp_\Aa(A)$ for all $A\in\Aa$.\footnote{This
follows using the fact that $\alpha(\Aa)$ is $C^*$-subalgebra of $\Ba$
(Prop. 2.3.1 of \cite{BratRob}) and because the spectrum of $\alpha(A)$,
relative to $\alpha(\Aa)$ is equal to its spectrum in $\Ba$ - Prop.
2.2.7 of \cite{BratRob}.} As with the numerical range, this entails
the existence of a natural mapping
$\Sp:\Af\nto 2^\CC$ expressed by commutativity of the diagram
\begin{equation*}
\begin{CD}
\Mb @.\phantom{AAAA}@.\Af(\Mb) @>\Sp_{\Mb}>> 2^\CC \\
@V{\psi}VV @. @V{\alpha(\psi)}VV     @VV{\id_{2^\CC}}V\\
\Nb @.\phantom{AAAA}@.\Af(\Nb)@>>\Sp_{\Nb}> 2^\CC
\end{CD}\qquad.
\end{equation*}
Hence, composing with any field $\Phi:\Df\nto\Af$, we obtain a natural map
$\Sp(\Phi):\Df\nto 2^\CC$ such that 
\begin{equation*}
\Sp(\Phi)_\Mb(f) = \Sp_{\Af(\Mb)}(\Phi_\Mb(f))\,.
\end{equation*}

In addition, we may also consider the spectrum of each
$\Phi\in\Fa^\infty(\Df,\Af)$, which we
denote $\sigma(\Phi)$ for short. In contrast to $\Sp(\Phi)$, this is not
a natural transformation, but simply a subset of $\CC$. Nonetheless,
there is a relation between the two.

\begin{Prop} \label{prop:sigma_and_Sp}
$\sigma(\Phi)$ and $\Sp(\Phi)$ are related by 
\begin{equation}
\sigma(\Phi) = \cl\left(\bigcup_{\Mb\in\Mfr}\bigcup_{f\in\Df(\Mb)}
\Sp(\Phi)_\Mb(f)\right)\,.
\label{eq:spec_eq}
\end{equation}
\end{Prop}
{\noindent\em Proof:} Suppose $\lambda\notin\sigma(\Phi)$. Then there
exists $\Psi\in\Fa^\infty(\Df,\Af)$ such that
\begin{equation*}
\Psi(\lambda\II-\Phi) =(\lambda\II-\Phi)\Psi = \II\,,
\end{equation*}
which entails that $\lambda$ belongs to the resolvent set of every
$\Phi_\Mb(f)$. Accordingly,
we see that the right-hand side of \eqref{eq:spec_eq} is contained in the
left (as $\sigma(\Phi)$ is closed). Conversely, if $\lambda$ does not
belong to the right-hand side of \eqref{eq:spec_eq}, then there exists
$\epsilon>0$ for which the disc $\{\mu\in\CC:|\mu-\lambda|<\epsilon\}$ lies in the
resolvent set of every $\Phi_\Mb(f)$ (initially for $\Mb\in\Mfr$, but 
hence for all $\Mb\in\Man$). Setting $\Psi_\Mb(f) =
(\lambda\II_{\Af(\Mb)}-\Phi_\Mb(f))^{-1}$, we have the uniform bound
\begin{equation}
\label{eq:uniform_bound}
\|\Psi_\Mb(f)\| \le \epsilon^{-1} \qquad \forall
\Mb\in\Man,~f\in\Df(\Mb)\,.
\end{equation}
If $\psi:\Mb\to\Nb$ in $\Man$, we apply $\alpha_\psi$ to the equation
$\Psi_\Mb(f) (\lambda\II_{\Af(\Mb)}-\Phi_\Mb(f)) = \II_{\Af(\Mb)} =
 (\lambda\II_{\Af(\Mb)}-\Phi_\Mb(f))\Psi_\Mb(f)$ to deduce that
$\alpha_\psi\Psi_\Mb(f) = \Psi_\Nb(\psi_*f)$, and hence that the
$\Psi_\Mb$ constitute a natural transformation $\Psi:\Df\to\Af$. The
bound \eqref{eq:uniform_bound} then entails that
$\Psi\in\Fa^\infty(\Df,\Af)$, so $\lambda\II-\Phi$ is invertible in $\Fa(\Df,\Af)$,
completing the proof. $\square$

Fields in $\Fa^\infty(\Df,\Af)$ may be manipulated according to
functional calculus: for example, if $\Phi\in\Fa^\infty(\Df,\Af)$ is
normal and $\varphi:\sigma(\Phi)\to\CC$ is continuous then there is an
element $\varphi(\Phi)\in\Fa^\infty(\Df,\Af)$, with
$\sigma(\varphi(\Phi))=\varphi(\sigma(\Phi))$. The field
$\varphi(\Phi)$ is automatically covariant, and obeys
$\varphi(\Phi)_\Mb(f) = \varphi(\Phi_\Mb(f))$. While the latter could
also serve as a definition, we would need to check naturality. The advantage of
using our algebra $\Fa^\infty(\Df,\Af)$ is that naturality is automatic, so
we have a manifestly covariant functional calculus. 

We also obtain a new definition of a positive field, as one whose
spectrum is positive, i.e., $\sigma(\Phi)\subset
[0,\infty)$. Standard $C^*$-algebra theory entails that
$\Phi\in\Fa^\infty(\Df,\Af)$ is positive if and only if it is the square of another
field; we also have that $\Phi^*\Phi$ is positive for any $\Phi$.

We may easily recover
one of the key properties of the numerical range, provided that $S$ is
sufficiently large. 

\begin{Lem} \label{lem:Sp_and_NR}
Suppose $\Aa$ is a $C^*$-algebra, and $S$ is weak-$*$
dense in $\Aa^*_{+,1}$. If $N_{\Aa,S}(A)$ is contained in the real axis
then the convex hull of the $\Sp_\Aa(A)$ is
\begin{equation*}
\co\,\Sp_\Aa(A) = N_{\Aa,S}(A)\,.
\end{equation*}
In particular, this holds if $S$ contains at least one state inducing
a faithful representation of $\Aa$
and is closed under operations induced by $\Aa$. 
\end{Lem}
{\noindent\em Proof:} Using weak-$*$ density and the fact that we have
required the numerical range to be closed, we have
$N_{\Aa,S}(A)=N_{\Aa,\Aa_{+,1}^*}(A)$. This is equal to the standard definition
of the numerical range of $A$ as in~\cite{BonsallDuncan,Palmer_vol1}
because the numerical range turns out to be closed for elements of
$C^*$-algebras (Proposition 2.6.2(a) in~\cite{Palmer_vol1}). The result then
follows using the standard result for numerical range, e.g., Theorem
2.6.7(d) in~\cite{Palmer_vol1}. The last
statement follows by the proof of Lemma~\ref{lem:Fell}. $\square$

\begin{Prop} Suppose each $\Af(\Mb)$ is a $C^*$-algebra and each
$\Sf(\Mb)$ is closed under operations induced by $\Af(\Mb)$ and contains 
at least one state inducing a faithful GNS representation of
$\Af(\Mb)$. If $\Phi\in\Fa^\infty(\Df,\Af)$ has numerical range
$\nu(\Phi)\subset\RR$ then
\begin{equation*}
\co\,\sigma(\Phi)= \nu(\Phi).
\end{equation*}
\end{Prop}
{\noindent\em Proof:} By Lemma~\ref{lem:Fell} and
Proposition~\ref{prop:LPE_from_Fell}, we know that
each $\Sf(\Mb)$ is weak-$*$ dense in $\Af(\Mb)^*_{+,1}$ and that $\Sf$
respects local physical equivalence. Hence we may define the numerical
range $N(\Phi):\DD\nto 2^\CC$. Now $\nu(\Phi)\subset
N_{\Fa^\infty(\Df,\Ff),\Fa^\infty(\Df,\Ff)^*_{+,1}}(\Phi)$ because it is a numerical
range over a subset of states. But the latter set is equal to
$\co\,\sigma(\Phi)$ by Lemma~\ref{lem:Sp_and_NR}, so
$\nu(\Phi)\subset\co\,\sigma(\Phi)$, and is, in particular, bounded. 

On the other hand, by Lemma~\ref{lem:Sp_and_NR} we have
\begin{equation*}
\Sp(\Phi)_\Mb(f) \subset \co\,\Sp_{\Af(\Mb)}(\Phi_\Mb(f)) =
N_{\Af(\Mb),\Sf(\Mb)}(\Phi_\Mb(f)) = N(\Phi)_\Mb(f)\,.
\end{equation*}
Taking the union
over all $\Mb\in\Mfr$ and $f$, closing, and forming the convex
hull, we obtain [also using Proposition~\ref{prop:sigma_and_Sp}]
\begin{equation*}
\co\,\sigma(\Phi)\subset \co\,\cl \left(\bigcup_{\Mb\in\Mfr}\bigcup_{f\in\Df(\Mb)}
N(\Phi)_\Mb(f)\right)=\cl\,\co\left(\bigcup_{\Mb\in\Mfr}\bigcup_{f\in\Df(\Mb)}
N(\Phi)_\Mb(f)\right)=\nu(\Phi)\,,
\end{equation*}
because $\cl$ and $\co$ commute on
bounded sets in $\RR^k$ (see e.g., Theorem 17.2
in~\cite{Rockafellar}, although this is essentially obvious in our
$1$-dimensional case). Hence
$\co\,\sigma(\Phi)=\nu(\Phi)$ as required.  $\square$

An interesting point about this result is that $S(\Df,\Af)$ contains
sufficiently many states to guarantee the usual connection between spectrum
and numerical range, even though we do not know whether it is weak-$*$
dense in $\Fa^\infty(\Df,\Af)^*_{+,1}$. 

\section{Conclusion}\label{sect:conclusion}

The main purpose of this paper has been to locate quantum (energy)
inequalities within the categorical framework of local covariance
developed by BFV. This has led us to a broader definition of QIs than
has previously been adopted, because we allow for the possibility of
state-dependent lower bounds. This seems natural from the
categorical point of view and also appears to be needed in some specific
instances, including the non-minimally coupled scalar
field~\cite{FO07}. We have also given a first attempt to delineate when
such a bound should be regarded as trivial, taking our inspiration from
the sharp G{\aa}rding inequalities, 
and we have checked that our definitions respect covariance and
are compatible with each other in various ways. In the process it has
become clear that the property of local physical equivalence, isolated
here for the first time, plays an important role in the analysis of
locally covariant quantum field theories. We have also considered the
broader question of the definition and basic properties of covariant numerical
range and spectrum of local quantum fields, leading naturally to the abstract
algebras of fields $\Fa(\Df,\Af)$ and $\Fa^\infty(\Df,\Af)$. As an
application of some of our ideas, we have shown how information about
spatially toroidal spacetimes can be used to infer properties of
quantum field theory on Minkowksi space. 

Our work raises several questions for future study. Can one give a
formal, locally covariant, definition of what it means for one field to
be of `lower order' than another? In Minkowksi space one could appeal to
$H$-bounds to provide a scale of fields, but in general one has no
global Hamiltonian. It is hoped that the definition sketched in
Sect.~\ref{sect:AQIs} can be developed to meet this end. More broadly, is the notion of
triviality studied here a sufficiently stringent definition? If not, can a more
refined version be found? This might well take the form of a
grading on the elements of algebras such as $\Ff(\Df,\Af)$. 
It is also necessary to investigate the local physical equivalence
property in the context of known models. One would also like to make a more
precise connection between QIs and the phase space properties of a
theory, perhaps establishing them as precise analogues of the
sharp G{\aa}rding inequalities. In turn, this raises the question of how
the phase space of the theory may be controlled in the locally covariant
setting. Above all, a key question is to determine what structural
features of a locally covariant quantum field theory are sufficient to
guarantee the existence of QEIs. It is hoped to return to these questions elsewhere.

{\noindent\em Acknowledgements:} The author thanks the Max Planck
Institute for Mathematics in the Natural Sciences, Leipzig, and the
Institute for Theoretical Physics, G\"ottingen, for
hospitality during the course of this work. Parts of this
work were also conducted during the programme ``Global problems in
Mathematical Relativity II'' at the Isaac Newton Institute, Cambridge,
and the author thanks the programme organisers and the Institute for its
hospitality.  It is also a pleasure to thank Rainer Verch and Bernard Kay
for useful discussions on local physical equivalence, Victoria Gould for
discussions regarding categories and sets, and Ko Sanders, Lutz Osterbrink
and the referees for
comments on the manuscript. 

\appendix
\section{The Wick square of the free scalar field}\label{appx:Wick_square}

As a concrete example, we briefly consider a
difference QI on the Wick square of the free scalar field of mass $m$
obtained by the methods of~\cite{Fews00}. 
In BFV, the free scalar field theory was explicitly expressed in terms
of a functor $\Af:\Man\to\TAlg$, and the free scalar field itself as a
natural transformation $\varphi:\Df\nto\Af$, with each $\Df(\Mb)$ equal to
the smooth compactly supported complex-valued functions on $\Mb$.
Hollands and Wald~\cite{Ho&Wa01} constructed enlarged algebras $\Wf(\Mb)$ that
represent algebras of Wick polynomials in $\varphi_\Mb$, and these
algebras may be
regarded as the images on objects of a functor $\Wf:\Man\to\TAlg$. Each
$\Af(\Mb)$ may be regarded as a sub-$*$-algebra of $\Wf(\Mb)$.  
The state space $\Sf$ associated with $\Wf$ is defined so that
$\Sf(\Mb)$ consists of all states on $\Wf(\Mb)$ whose two-point
functions obey the Hadamard condition (expressed as a certain condition
on the wave-front set~\cite{Hormander1}), and whose truncated
$n$-point functions for $n\not = 2$ are all smooth. The
reader is referred to BFV and~\cite{Ho&Wa01,HollandsRuan} for more
details; below, we restrict attention to the portion of the structure
relevant to our discussion. 

Let $\EE'_1$ be the functor
$\EE'_1:\Man\to\Set$ defined to act on any object $\Mb\in\Man$ so that
\begin{equation*}
\EE'_1(\Mb)=\{f\in\EE'(\Mb): \WF(f)\cap \Vc_\Mb=\emptyset\}\,,
\end{equation*}
where $\EE'(\Mb)$ is the usual space of compactly supported (scalar)
distributions on $\Mb$, $\WF(f)$ denotes the wave-front set of
a distribution $f$ and $\Vc_\Mb\subset T^*\Mb$ is the bundle of causal covectors on $\Mb$.
Given any $\psi:\Mb\to\Nb$, we define $\psi_*$ to be the induced push-forward of
compactly supported distributions $\psi_*:\EE'(\Mb)\to\EE'(\Nb)$;
setting $\EE'_1(\psi)=\psi_*$ (or more precisely its restriction to
$\EE'_1(\Mb)$) it is easy to verify that $\EE'_1$ is a functor as
claimed, of which $\Df(\Mb)$ is a subfunctor. 

The free scalar field $\varphi$ extends to a natural transformation between
$\EE'_1$ and $\Wf(\Mb)$; by contrast, the Wick square is not uniquely
defined. Rather, there is a family of natural transformations between $\EE'_1$ and
$\Wf$ with the property that, if $\varphi^2$ and $\widetilde{\varphi}^2$ are any two members of the
family, then there are real constants $c_1$ and $c_2$ such that
\begin{equation*}
\varphi^2_\Mb(f)-\widetilde{\varphi}^2_\Mb(f) = f(c_1 R_\Mb+c_2 m^2)\II_{\Wf(\Mb)}
\end{equation*}
for all $f\in\EE'_1(\Mb)$, and each $\Mb\in\Man$. Here $R_\Mb$ is the
Ricci scalar on $\Mb$. Any one of these
natural transformations provides a valid definition of the Wick square:
we henceforth suppose that one has been chosen, which we denote
$\varphi^2$. (For a proposal to fix the renormalisation constants on
thermodynamic grounds, see~\cite{BuchSchlem07}). Given any $\omega,\omega'\in\Sf(\Mb)$, we also have
\begin{equation}
\omega(\varphi^2_\Mb(f))-\omega'(\varphi^2_\Mb(f))=\delta_2^*(\Lambda_\omega-\Lambda_{\omega'})(f)\,,
\label{eq:Waldax}
\end{equation}
where $\delta_2:\Mb\to\Mb\times\Mb$ is defined by $\delta_2(p)=(p,p)$
and $\Lambda_\omega$ denotes the two-point function of $\omega$, i.e.,
$\Lambda_\omega(p,p')=\omega(\varphi_\Mb(p)\varphi_\Mb(p'))$ in
unsmeared notation; $\delta_2^*$ denotes the distributional pull-back by $\delta_2$. 

Classically, of course, a squared field is pointwise nonnegative. In the 
quantum field theory, however, the quantised Wick square is capable of assuming negative
values, which turn out to be constrained by a DQI. For each
$\Mb\in\Man$, let $\Ff(\Mb)$ be the set of $f\in\EE_1'(\Mb)$ such that
\begin{equation*}
f(u) = \int_\gamma g(\tau)^2u(\gamma(\tau))\,d\tau \qquad (u\in C^\infty(\Mb))\,,
\end{equation*}
where $I$ is an open interval of $\RR$,
$\gamma:I\to \Mb$ is a proper time parameterisation of a smooth,
future-pointing timelike curve, and $g$ is a smooth 
real-valued function, compactly supported in $I$ and with no zeros of
infinite order in the interior of its support. Given $f$, we may
reconstruct $I$, $\gamma$ and $g$ up to reparameterisations which may be
ignored in the following discussion (see~\cite{Few&Pfen06} for details).
The DQI obtained by the methods of~\cite{Fews00} is
\begin{equation}
\omega(\varphi_\Mb^2(f))-\omega_0(\varphi_\Mb^2(f))\ge
-\int_0^\infty \frac{d\alpha}{\pi} \left[(g\otimes g)
\gamma_2^*\Lambda_{\omega_0}\right]^{\wedge}(-\alpha,\alpha)\,\,\stack{\rm def}{=}\,\,
-\widetilde{\QQ}_\Mb(f,\omega_0)\,,
\label{eq:DQI_F00}
\end{equation}
where $\gamma_2^*$ denotes the distributional pull-back from $M\times M$
to $\RR^2$ by the map
$\gamma_2(\tau,\tau')=(\gamma(\tau),\gamma(\tau'))$ and the hat denotes
a Fourier transform (see~\cite{Few&Pfen06} for the conventions).
In~\cite{Few&Pfen06} arguments are given which show that the central
member of \eqref{eq:DQI_F00} is
independent of the particular parameterisation of $f$ in terms of $I$,
$\gamma$, and $g$; these arguments also show that $\Ff$ is a subfunctor
of $\EE_1'$ and that the covariance relation
$\widetilde{\QQ}_\Nb(\psi_*f,\omega_0) =
\widetilde{\QQ}_\Mb(f,\psi^*\omega_0)$ holds. Thus $\QQ_\Mb(f,\omega_0)=
\widetilde{\QQ}_\Mb(f,\omega_0)\II_{\Af(\Mb)}$ defines a natural map
$\QQ:\langle\Ff,\Sf^\op\rangle\nto\Wf$, establishing this bound as a locally covariant
difference quantum inequality with respect to $\Sf$. 

As well as providing a concrete example of the various functors and
natural transformations involved in our framework, we also want to point
out two features of this bound not discussed in~\cite{Few&Pfen06}.
First, we prove that it satisfies the condition~\eqref{eq:independence}.
This is essentially the reverse of the argument in~\cite{Fews00}: noting
that 
\begin{equation*}
\widetilde{\QQ}_\Mb(f,\omega)-\widetilde{\QQ}_\Mb(f,\omega_0) =
\int_0^\infty \frac{d\alpha}{\pi} \left[(g\otimes g)
\gamma_2^*(\Lambda_{\omega}-\Lambda_{\omega_0})\right]^{\wedge}(-\alpha,\alpha)\,,
\end{equation*}
we use the fact that $\Lambda_{\omega}-\Lambda_{\omega_0}$ is symmetric
and smooth to rewrite the right-hand side as
\begin{eqnarray*}
\int_{-\infty}^\infty \frac{d\alpha}{2\pi} \left[(g\otimes g)
\gamma_2^*(\Lambda_{\omega}-\Lambda_{\omega_0})\right]^{\wedge}(-\alpha,\alpha)
&=& \int d\tau\, g(\tau)^2
\left(\Lambda_{\omega}-\Lambda_{\omega_0}\right)(\gamma_2(\tau))\\
&=& \omega(\varphi_\Mb^2(f))-\omega_0(\varphi_\Mb^2(f))
\end{eqnarray*}
as required, using the Fourier representation of the $\delta$-function
and \eqref{eq:Waldax}. Accordingly, as shown in Sect.~\ref{sect:DQIs},
$\QQ_\Mb(f)=\QQ_\Mb(f,\omega_0)-\omega_0(\varphi^2_\Mb(f))\II_{\Af(\Mb)}$
defines a locally covariant AQI, because the right-hand side is independent of
$\omega_0\in\Sf(\Mb)$. 

The second fact about our original DQI follows immediately from the
first: namely the map $\omega_0\mapsto \QQ_\Mb(f,\omega_0)$ is weak-$*$
continuous for each $f\in\Ff(\Mb)$; furthermore, $(\omega_0,\omega)\mapsto
\omega(\QQ_\Mb(f,\omega_0))$ is weak-$*$ continuous in $\omega_0$,
uniformly in $\omega\in\Sf(\Mb)$. This shows that this DQI obeys the
continuity hypotheses required in
Propositions~\ref{prop:DQI_to_AQI_by_inf} and~\ref{prop:covariance_of_triviality}(b).

\section{The non-minimally coupled scalar field}\label{appx:nonminimal}

Recall that the non-minimally coupled field obeys the field equation $(\Box+m^2+\xi
R)\varphi=0$, where $\xi$ is called the coupling constant; its stress-energy
tensor differs from that of the minimally coupled ($\xi=0$) theory by
terms which permit the violation of the pointwise energy conditions even
at the classical level. One may show that this renders it impossible to
obtain state-independent QEIs~\cite{FO07}. Our purpose in this Appendix
is to illustrate the discussion of state-dependent QEIs with an example
to show the type of state-dependence that enters in the difference QEI
obtained in \cite{FO07} by the present author and Osterbrink for the
case $0\le\xi\le 1/4$.

Let $\EE_1'T_0^2:\Man\to\Set$ be the functor assigning to each
$\Mb\in\Man$ the set of compactly supported rank-$2$ contravariant
tensor distributions
\begin{equation*}
\EE_1'T_0^2(\Mb) = \{f\in \EE'(T_0^2\Mb):~\WF(f)\cap\Vc_\Mb=\emptyset\}\,,
\end{equation*}
with $\Vc_\Mb$ as in Appendix~\ref{appx:Wick_square}, and $\EE_1'T_0^2$
acting on morphisms to give the corresponding push-forward. Then the
stress-energy tensor of the theory may be regarded as a natural
transformation $T:\EE_1'T_0^2\nto\Wf$, using the usual point-splitting
prescriptions to define $T_\Mb$ in any particular spacetime $\Mb$. (As
with the Wick square, there is a family of possible different
candidates, and we assume that one has been chosen.) We
define a subfunctor $\Ff\subset \EE_1'T_0^2$ so that $\Ff(\Mb)$ consists
of those $f\in\EE_1'T_0^2(\Mb)$ of the form
\begin{equation}
f(t) = \int_\gamma g(\tau)^2\dot{\gamma}^a\dot{\gamma}^b t_{ab}|_{\gamma(\tau)}\,d\tau \qquad (t\in C^\infty(T_2^0\Mb))\,,
\label{eq:smearing_def}
\end{equation}
where $\gamma:I\to \Mb$ is a proper time parameterisation of a smooth, future-directed
timelike {\em geodesic}, with $I\subset \RR$ open, and $g\in\CoinX{I;\RR}$
obeying the same hypotheses as in Appendix~\ref{appx:Wick_square}. The QEI
obtained in~\cite{FO07} can be expressed in the form
\begin{equation*}
\omega(T_\Mb(f))-\omega_0(T_\Mb(f))\ge -
\omega(\QQ_\Mb(f,\omega_0))\,,
\end{equation*}
for states $\omega,\omega_0\in\Sf(\Mb)$ and any $f\in\Ff(\Mb)$, where
$\QQ:\langle\Ff,\Sf^\op\rangle\nto\Wf$ is natural and has components
$\QQ_\Mb(f)\in\Wf(\Mb)$ of the form
\begin{equation*}
\QQ_\Mb(f,\omega_0) = \widetilde{\QQ}_\Mb(f,\omega_0)\II_{\Wf(\Mb)} +
\varphi_\Mb^2(S_\Mb(f))\,,
\end{equation*}
where $\varphi^2$ is the renormalised Wick square. Here, $S_\Mb$ is a
component of a natrual transformation $S:\Ff\nto\EE_1'$,
defined so that
\begin{equation}
(S_\Mb(f))(u) = f(u W_\Mb)+ 2\xi \int_\gamma 
\dot{g}(\tau)^2 u(\gamma(\tau))\,d\tau \qquad (u\in C^\infty(\Mb))\,,
\end{equation}
for $f$ of the form \eqref{eq:smearing_def} and the tensor $W_\Mb$
multiplying $u$ is given by
\begin{equation*}
W_\Mb = \xi \text{Ric}_\Mb - \frac{1}{2}\xi(1-4\xi)R_\Mb g_\Mb
\end{equation*}
where $\text{Ric}_\Mb$, $R_\Mb$ and $g_\Mb$ are the Ricci tensor, Ricci
scalar and metric of $\Mb$. Note that $W_\Mb$ vanishes if $\xi=0$ or $\Mb$ is Ricci
flat, and $S_\Mb(f)=0$ for any $f$ if $\xi=0$, thus rendering the bound
state-independent in this case. 
[In the above, $\varphi^2$, $\Wf$, $\Sf$ and $\EE_1'$ are all as in
Appendix~\ref{appx:Wick_square}.] In addition, $\widetilde{\QQ}_\Mb(f)$ also depends on $\xi$, and reduces
to a previously known state-independent bound if $\xi=0$. Details appear in \cite{FO07}. 
For our current purpose the main point is that the state-dependence is a
smearing of the Wick square $\varphi^2$, whereas the stress-energy
tensor itself involves squares of derivatives of $\varphi$. This fits
well with the analogy between nontrivial QEIs and the gain in derivatives
exhibited by the G{\aa}rding inequalities. In~\cite{FO07} we also show that the
bound is nontrivial in the sense described in this paper: for example,
in thermal states on Minkowski space the lower bound scales like the
square of the temperature while the smeared stress-energy tensor itself
scales with the fourth power.


\end{document}